\PassOptionsToPackage{table}{xcolor}
\documentclass[sigconf,screen,nonacm]{acmart}
\usepackage{url}

\usepackage{amsmath,amssymb,amsfonts}
\usepackage{tabularx}
\usepackage{subcaption}
\usepackage{algorithmic}
\usepackage{graphicx}
\usepackage{textcomp}
\usepackage{multirow}
\usepackage{booktabs}
\usepackage{tabularx}
\usepackage{tikz}
\usepackage{enumitem}

\AtBeginDocument{%
  }

\setcopyright{acmlicensed}
\copyrightyear{2018}
\acmYear{2018}
\acmDOI{XXXXXXX.XXXXXXX}
\acmConference[ICSE '26]{Make sure to enter the correct
  conference title from your rights confirmation email}{April 12--18,
  2026}{Rio de Janeiro, Brazil}
\acmISBN{978-1-4503-XXXX-X/2018/06}

\setcopyright{none}
\renewcommand\footnotetextcopyrightpermission[1]{} 

\begin{document}

\title{Improving Compiler Bug Isolation by Leveraging Large Language Models}


\author{Yixian Qi}
\affiliation{%
  \institution{Tianjin University}
  \city{Tianjin}
  \country{China}}
\email{qiyixian@tju.edu.cn}

\author{Jiajun Jiang}
\affiliation{%
  \institution{Tianjin University}
  \city{Tianjin}
  \country{China}}
\email{jiangjiajun@tju.edu.cn}

\author{Fengjie Li}
\affiliation{%
  \institution{Tianjin University}
  \city{Tianjin}
  \country{China}}
\email{fengjie@tju.edu.cn}

\author{Bowen Chen}
\affiliation{%
  \institution{Tianjin University}
  \city{Tianjin}
  \country{China}}
\email{3022244160@tju.edu.cn}

\author{Hongyu Zhang}
\affiliation{%
  \institution{Chongqing University}
  \city{Chongqing}
  \country{China}}
\email{hongyujohn@gmail.com}

\author{Junjie Chen}
\affiliation{%
  \institution{Tianjin University}
  \city{Tianjin}
  \country{China}}
\email{junjiechen@tju.edu.cn}







\renewcommand{\shortauthors}{Qi et al.}



\newcommand{\tool}[1]{\textsc{AutoCBI}}
\newcommand{\variant}[1]{\tool{}$_{#1}$}

\newcommand{\jiajun}[1]{{\color{cyan}[Jiajun: #1]}}
\newcommand{\hy}[1]{{\color{red}[HY: #1]}}
\newcommand{\jj}[1]{{\color{orange}[Junjie: #1]}}
\newcommand{\yixian}[1]{{\color{blue}[Yixian: #1]}}

\newcommand*\encircle[1]{\tikz[baseline=(char.base)]{\node[shape=circle,draw,fill=black,text=white,inner sep=0.5pt] (char) {#1};}}

\newcommand{\distance}{3pt}
\setlength{\textfloatsep}{\distance}
\setlength{\floatsep}{\distance}
\setlength{\dbltextfloatsep}{\distance} 
\setlength{\dblfloatsep}{\distance} 

\begin{abstract}

Compilers play a foundational role in building reliable software systems, and bugs within them can lead to catastrophic consequences. The compilation process typically involves hundreds of files, making traditional automated bug isolation techniques inapplicable due to scalability or effectiveness issues. Current mainstream compiler bug localization techniques have limitations in test program mutation and resource consumption.
Inspired by the recent advances of pre-trained Large Language Models (LLMs), 
we propose an innovative approach named \tool{}, which (1) uses LLMs to summarize compiler file functions and (2) employs specialized prompts to guide LLM 
in reordering suspicious file rankings. This approach leverages four types of information: 
the failing test program, source file function summaries, lists of suspicious files identified through analyzing test coverage, as well as compilation configurations with related output messages, resulting in a refined ranking of suspicious files.
Our evaluation of \tool{} against state-of-the-art approaches (DiWi, RecBi and FuseFL) on 120 real-world bugs from the widely-used GCC and LLVM compilers demonstrates its effectiveness. Specifically, \tool{} isolates 66.67\%/69.23\%, 300\%/340\%, and 100\%/57.14\% more bugs than RecBi, DiWi, and FuseFL, respectively, in the Top-1 ranked results for GCC/LLVM. Additionally, the ablation study underscores the significance of each component in our approach.
\end{abstract}
\maketitle

\section{Introduction}
\label{sec:intro}
Compiler bugs present critical challenges in software systems, potentially leading to catastrophic consequences~\cite{chen2020survey} such as system crashes~\cite{sun2016finding}, erroneous program behaviors~\cite{wang2013towards}, and security vulnerabilities~\cite{d2015correctness,sidhpurwala2019security}. Ensuring the robustness and reliability of compilers is crucial, as they are foundational tools that take the responsibility to translate high-level source code into low-level executable programs. Compiler bugs may propagate to application programs compiled by the compilers due to incorrect translations, which will seriously aggravate the difficulty of the downstream program debugging process since it is hard to determine whether the unexpected behaviors in application programs are from the used compilers or the program themselves. Consequently, the identification and debugging of compiler bugs are vital tasks in maintaining software integrity and reliability.

To tackle this issue, a variety of compiler testing methods have been proposed and developed in the literature~\cite{abreu2007accuracy,digiuseppe2011influence,zhang2011localizing,li2017transforming,wen2019historical}, aiming at uncovering potential bugs by generating diverse and comprehensive test cases. These approaches leverage techniques such as random testing~\cite{duran1984evaluation}, fuzz testing\cite{godefroid2008automated} and deep learning~\cite{lecun2015deep}, which have shown effectiveness in discovering subtle and critical flaws in compilers. Empirical studies have demonstrated the efficacy of these methods in identifying bugs across widely used compilers (i.e., GCC~\cite{gcc} and LLVM~\cite{llvm}). However, despite these advancements, the debugging process following the identification of test failures remains time-consuming and labor-intensive. The challenge is underpinned by two primary factors: First, due to the inherent complexity of compilers, failing test programs often involve numerous interdependent files, all of which are potential sources of failures. As a result, the intricate interrelation of compiler components complicates the isolation of the root cause, often requiring significant manual effort to localize the actual responsible source files of the failure. Second, existing test programs are frequently too weak to distinctly separate faulty files from none-faulty ones, resulting in reduced effectiveness of existing coverage-based fault localization techniques~\cite{zhang2011localizing,xuan2014test,zhang2013injecting,lou2020can,benton2020effectiveness}. 

To tackle these challenges, existing compiler bug isolation approaches enhance test coverage differentiation through two main techniques. The first technique involves mutating compiler configurations to achieve varied coverage of compiler files, aiding in the identification of bugs~\cite{zhou2022locseq,yang2022isolating}. However, this method can be too coarse-grained, as an individual optimization may interact with multiple source files, complicating the accurate pinpointing of specific buggy locations. The second technique focuses on generating additional passing test programs that share similar execution paths with failing tests, enabling the isolation of faulty files through coverage analysis~\cite{DiWi,RecBi,llm4cbi}. Nonetheless, the complexities of compiler programs and their intricate execution path constraints make it difficult to produce effective passing test cases with limited mutations~\cite{DiWi,RecBi}. Although these methods are effective in isolating compiler bugs, they remain constrained by their dependence on test coverage alone.

Additionally, existing approaches often fail to fully utilize the insights developers gain during manual debugging, such as the characteristics of failing test programs, error messages, and compiler semantics. Developers typically draw on their deep understanding of code functions, synthesizing information from various sources to correlate test features, observed failures, and relevant code components~\cite{jiang2019manual}. This informed approach helps narrow the search space, allowing for more effective identification and resolution of bugs. Unfortunately, current compiler bug isolation methods overlooked these insights, which may hinder their effectiveness. While information retrieval (IR)-based techniques~\cite{6227210} have proven successful in fault localization, their reliance on high-quality bug reports~\cite{wang2015evaluating}, which are often lacking for compiler bugs~\cite{romano2021empirical}, limits their applicability and effectiveness in this context.

Inspired by the recent advances of large language models (LLMs), which have demonstrated remarkable effectiveness in various software engineering~\cite{jin2023inferfix,zhang2024autocoderover,hou2023large} and natural language processing tasks~\cite{min2023recent}, this paper proposes an innovative compiler bug isolation approach, named \tool{}, which aims at improving the compiler bug isolation performance by leveraging LLMs' capabilities in both natural language and code understanding for integrating additional contextual information into the debugging process. This integration provides us a new way to significantly improve the performance of compiler bug isolation by considering a broader spectrum of data, including both code and documents, which are critical for accurate diagnosis. Specifically, \tool{} currently integrates four types of information, i.e.,
the failing test program,
function summaries of implicated source files , lists of suspicious files identified through analyzing test coverage, as well as compilation
configurations and related output messages
with a specially designed prompt. 

To assess the effectiveness of \tool{}, we have conducted an extensive experiment using a widely-used dataset consisting of 120 real-world reproducible bugs from two prominent compilers, i.e., GCC~\cite{gcc} and LLVM~\cite{llvm}. We compare \tool{} with state-of-the-art baselines. The experimental results demonstrated that \tool{} successfully localized the faulty files within the top 1/5/10/20 positions for 42/63/78/95 bugs, significantly outperforming the best-performing baseline (i.e., RecBi) with improvements of at least 68.00\%/21.15\%/9.86\%/6.74\%, thereby highlighting the effectiveness of our approach. Additionally, our ablation study indicated that each type of information utilized in \tool{} significantly contributes to its effectiveness. Interestingly, the results show that not all types of information are necessary for localizing certain bugs, and in some cases, the inclusion of certain information can even negatively impact the localization performance. This finding suggests the need for further investigation and the development of adaptive methods for combining various information optimally. 

In summary, this work makes the following major contributions:
\begin{itemize}
    \item This work provides a new way of improving compiler bug isolation by integrating data from multiple sources beyond the coverage information of test execution. 
    \item This work introduces the first compiler bug isolation method utilizing LLM to incorporate various context features (named \tool{}). 
    \item We have conducted extensive experiments to evaluate the performance of our method against baseline approaches, highlighting its superiority and the significance of each component within our method.
    \item We have open-sourced all our implementations and experimental results to facilitate replication and support future research. \url{https://anonymous.4open.science/r/AutoCBI-D823/} 
\end{itemize}

\section{Motivation}
\label{sec:motivate}

\begin{figure}[tbp]
\centering
\begin{subfigure}{0.46\columnwidth}
    \includegraphics[width=\textwidth]{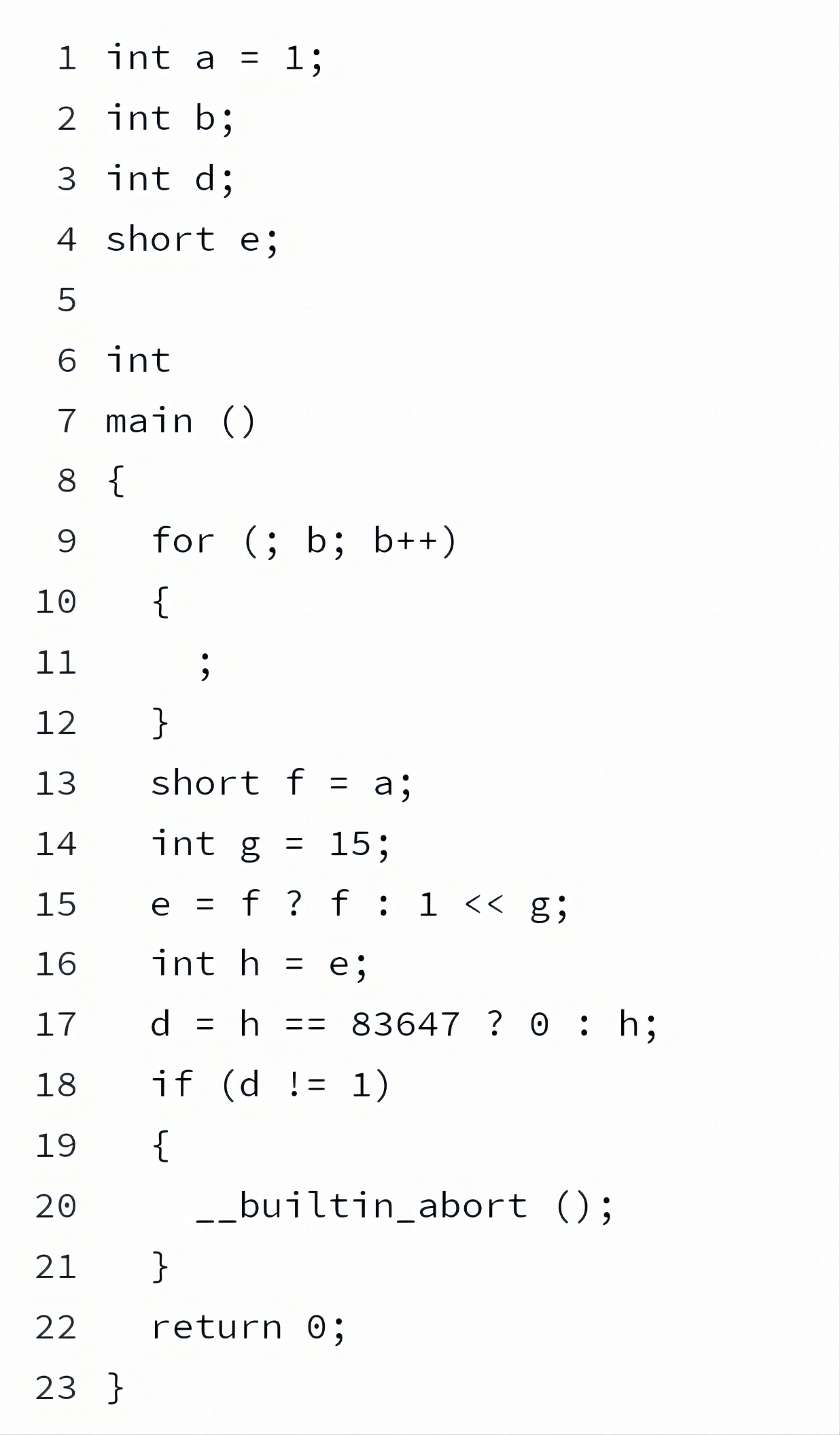}
    \caption{Failing test program.}
    \label{fail}
\end{subfigure}
\hspace{10pt}
\begin{subfigure}{0.42\columnwidth}
    \includegraphics[width=\textwidth]{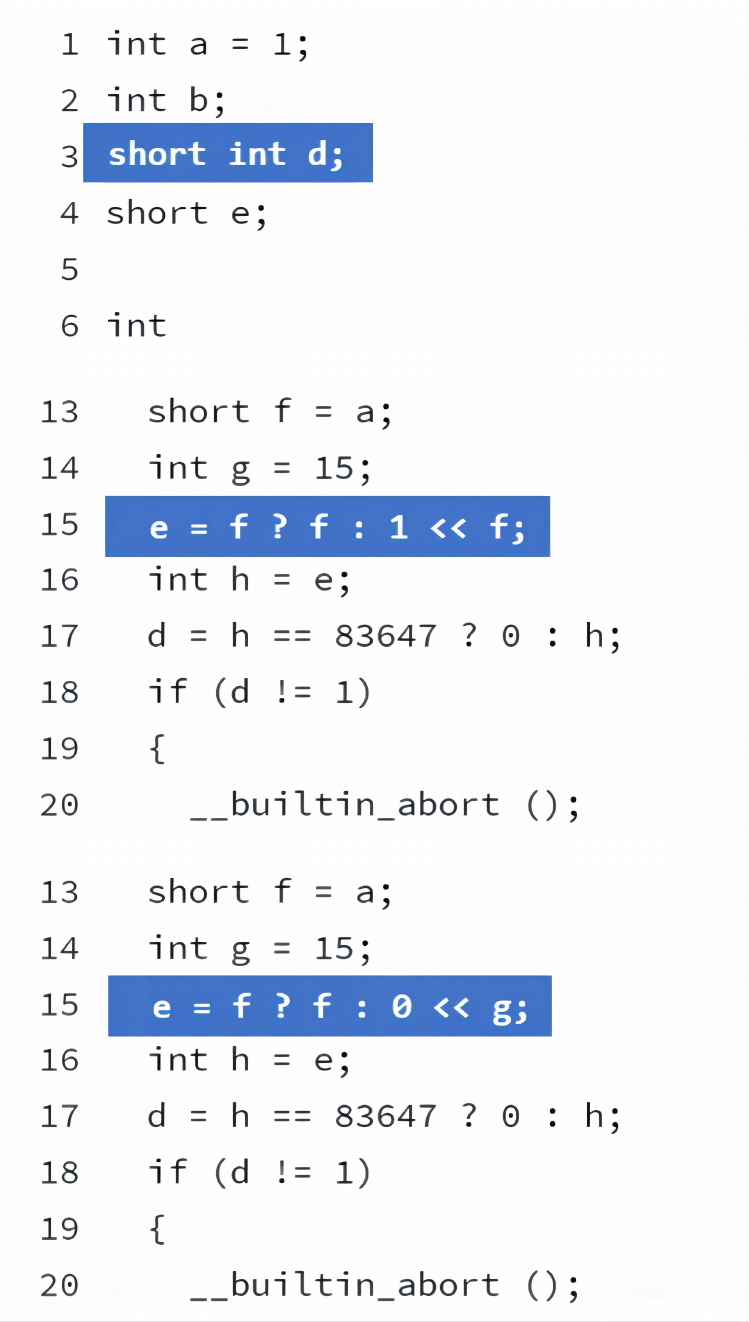}
    \caption{Three passing test programs generated by RecBi.}
    \label{pass}
\end{subfigure}
\caption{The failing test program triggering GCC bug (\href{https://gcc.gnu.org/bugzilla/show_bug.cgi?id=59221}{ID: 59221~\cite{GCCBUGID59221}}) and passing test programs by mutating the failing one. The mutated lines are highlighted.}
 \label{fig1}
\end{figure}

In this section, we use an example to illustrate the limitations of existing compiler bug isolation methods and motivate our approach. Fig.~\ref{fig1} presents an example from the GCC compiler (\href{https://gcc.gnu.org/bugzilla/show_bug.cgi?id=59221}{Bug ID: 59221~\cite{GCCBUGID59221}}), including the reported bug-triggering test program. The bug is located in the source file of ``{\tt tree-ssa-threadupdate.c}''. In this case, when the compiler (revision 205097 in GCC trunk) compiles the test program, an issue with the Value Range Propagation (VRP) optimization leads to incorrect code generation at optimization levels {\tt -O2} and {\tt -O3} on x86\_64-linux-gnu systems. 
As introduced, existing approaches perform bug isolation by either mutating failing test programs to generate passing ones or mutating the compiling optimization configurations.

\textbf{Bug isolation by mutating test programs:} 
The complexity of compiler programs makes generating suitable passing test programs challenging, undermining fault localization effectiveness. For example, in this example (GCC bug \href{https://gcc.gnu.org/bugzilla/show_bug.cgi?id=59221}{ID: 59221~\cite{GCCBUGID59221}}), 277 source files are involved in compiling the failing test program. To isolate the faulty file, passing programs should cover the other 276 files, but intricate dependencies between them make this almost impossible. Mutated test programs are often too similar to the failing one, covering the same files and leading to identical execution failures. Specifically, RecBi~\cite{RecBi}, a state-of-the-art compiler bug isolation method, generated only three valid passing test programs in an hour (see Fig.~\ref{pass}, partial test programs with mutations marked in blue). Existing methods typically produce very few passing test programs within a time limit, limiting their effectiveness and often resulting in inaccurate identification of buggy files, as shown by this example where the faulty file ranked outside the top 20 positions.

\textbf{Bug isolation by mutating optimizations:} 
Optimization configurations determine which components of a compiler execute during the compilation of a test program. Existing approaches modify these options or their order in an effort to make a failing test program pass, bypassing the faulty files. As a result, files not covered by successful test programs become likely candidates for faults. However, inherent dependencies among source files complicate this bypassing. When successful, significant coverage differences typically emerge between failing and passing executions, making accurate isolation challenging. For instance, in the example shown in Fig.~\ref{fig1}, bypassing optimizations related to VRP to make the failing test program pass involves multiple source files, any of which could be faulty.

Consequently, the performance of existing approaches remains limited due to the inherent complexity of compiler bug isolation. Importantly, these limitations are not isolated instances. Previous evaluations~\cite{RecBi,zhou2022locseq,yang2022isolating} indicate that state-of-the-art methods from these two categories can accurately localize at most 22.50\% and 24.16\% of compiler bugs within the top 1 position. Given thousands of bug reports related to GCC and LLVM compilers, there is an urgent need to 
develop more effective methods for compiler bug isolation due to the fundamental role of compilers.


Consider the typical debugging process undertaken by developers, who often draw upon a deep understanding of test failures and the functions of the involved code components. In these situations, developers may analyze the error messages
generated by the test program and the characteristics exhibited by the failing test. This analysis guides them toward identifying the components that might produce such messages or are responsible for handling the corresponding code features. For the example depicted in Fig.~\ref{fig1}, the control flow structure (associated with the {\tt for} loop) causes the failure. 
Specifically, during Value Range Propagation (VRP) optimization at the {\tt -O2} and {\tt -O3} optimization levels, the VRP process mishandles the control flow pattern introduced by the {\tt for} loop, leading to a compilation crash.
By checking the code semantics developers may understand that the source file ``{\tt tree-ssa-threadupdate.c}'' is closely related to the control flow updating as described by ``\textit{Thread edges through blocks and update the control flow and SSA graphs}'' in the corresponding document of the source file. As a result, developers may have more confidence to believe that this file has a high possibility to be the root cause. However, existing approaches have never utilized such information for compiler bug isolation. Nevertheless, accurately understanding the code features causing the compilation failure and further correlating the failure with the source file are indeed challenging.
To the best of our knowledge, we are the first to explore the capability of further incorporating such 
external information for improving compiler bug isolation by leveraging LLMs. Actually, our approach complements existing compiler bug isolation methods and provides a general framework that can combine existing methods (offering extra features) for further enhancement. As a result, the faulty file for the GCC compiler bug (\href{https://gcc.gnu.org/bugzilla/show_bug.cgi?id=59221}{ID: 59221}) is accurately localized at top 1 by our approach.

\section{The Design of \tool{}}
\label{sec:approach}
\begin{figure}[tbp]
\centerline{\includegraphics[width=1 \linewidth]{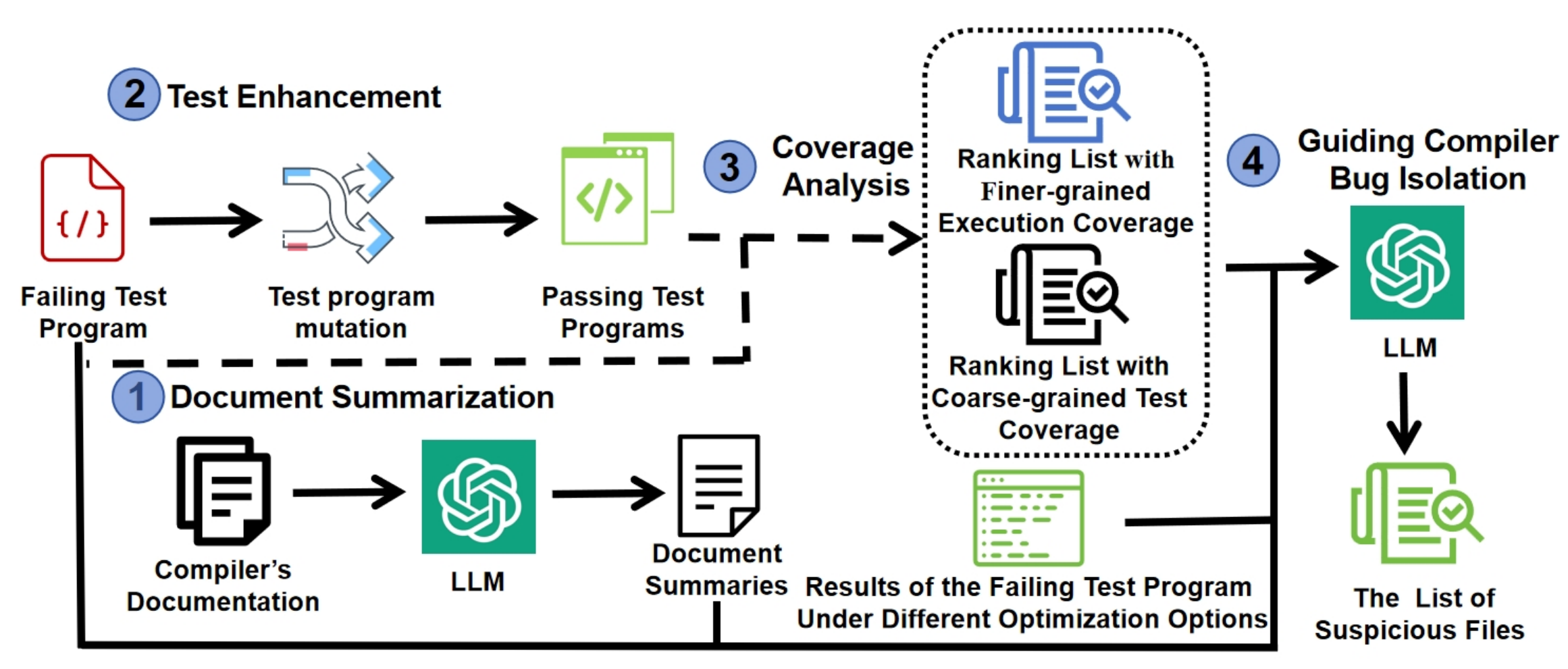}}
\caption{The framework of \tool{}}
\label{fig4}
\end{figure}
This section outlines the detailed processes in our workflow and the design of the prompts, which aim to optimize compiler bug localization by integrating multi-source information with the capabilities of LLMs in both natural language and code understanding.
Fig.~\ref{fig4} presents the overall workflow of our tool, \tool{} (\textbf{Auto}mated \textbf{C}ompiler \textbf{B}ug \textbf{I}solation), encompassing four core components. \encircle{1} Document summarization -- converts rich-text documents of source code into concise summaries, preserving key insights about code function to minimize noise. \encircle{2} Test enhancement -- generates a set of passing test programs to improve the distinguishing ability of test coverage; \encircle{3} Coverage analysis -- effectively utilizes test coverage features to enhance fault localization; \encircle{4} Prompt engineering -- develops specialized LLM prompts to integrate multi-source data  for guiding compiler bug isolation. In the following, we will delve into the details of each component.


\subsection{Document Summarization}
\label{sec:summary}

As highlighted in Section~\ref{sec:motivate}, understanding the function of source code is crucial for developers to identify the root causes of test failures during debugging. Documents accompanying source files are pivotal in this comprehension process. Therefore, effectively leveraging these documents to enhance function understanding and improve bug isolation performance is essential.

\begin{figure}[tbp]
\centerline{\includegraphics[width=1\columnwidth]{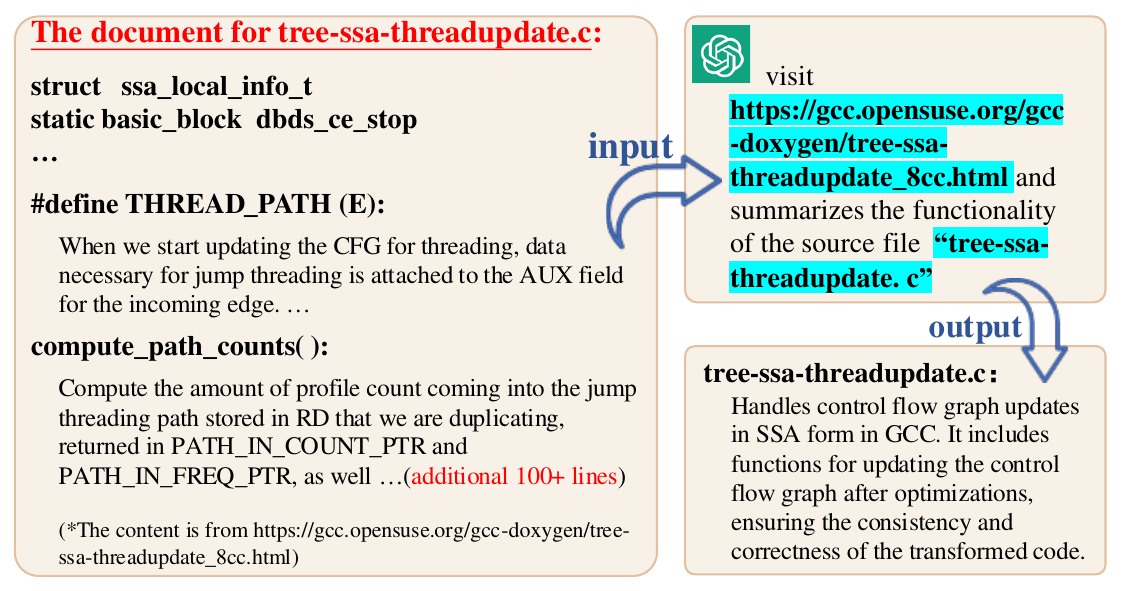}}
\caption{The simplified document for the source file ``{\tt tree-ssa-threadupdate.c}'' from the GCC compiler (on the left) and its summary generated by ChatGPT-4o along with the used prompt (on the right).}
\label{fig5}
\end{figure}

However, automatically interpreting these documents is challenging, as they often include rich-text formats with natural language and code snippets, leading to lengthy descriptions. Given the constraints on LLM input length and the need to integrate multiple sources, it’s vital to refine these documents before processing them with LLMs. To address this, \tool{} implements a preprocessing step to refine the associated source code documents prior to bug isolation. Specifically, \tool{} leverages LLMs to convert the rich-text documents into summaries. For example, Fig.~\ref{fig5} shows the document excerpt of the source file ``{\tt tree-ssa-threadupdate.c}'', where the faulty code reside for the GCC compiler bug triggered by the failing test program shown in Fig.~\ref{fail}. Specifically, the document often comprises two major information: (1) \textit{source code}, including the declarations of header files,  the definitions of data structures, macros, methods, and variables; (2) \textit{natural language}, explaining the function of each method defined in the file. Consequently, 
directly feeding such rich-text formatted documents into LLMs, along with other data sources, can potentially exceed their maximum input length and introduce noise, hindering program semantic understanding. This underscores the need for a document summarization component in \tool{}. This component aims to generate precise summaries for each source file by emphasizing core insights and omitting irrelevant details.

Notably, this is a one-time process; the generated summaries can be reused unless the document is updated. Fig.~\ref{fig5} illustrates the prompt designed for producing these document summaries using LLMs (The text highlighted in the prompt will be replaced for different source files). Specifically, \tool{} leverages ChatGPT-4o for generating document summaries, as it can automatically access a given website link and interpret its contents. 
More specifically, we first used a crawler to collect the document links to all the source files for the specified compilers from their official websites. Subsequently, we established mapping relationships between the files and their corresponding links, which will be supplied to the LLM for summary generation.
In this figure, we also present the resulting summary for the given document (shown in the figure), demonstrating how it effectively captures the core function of the source file in natural language by removing extraneous information, such as code definitions. 
As explained in Section~\ref{sec:motivate}, the test failure is closely correlated with the control structure (i.e., the {\tt for} loop), which is mishandled during performing the value range propagation optimization, and thus produce compilation failures. It is evident that the summary produced by the LLM shown in the figure can effectively facilitate the understanding of the code function, and thus improve the localization performance. In our experiment, the desired source file ``{\tt tree-ssa-threadupdate.c}'' was successfully localized at the top 1 position by our approach, reflecting the importance of the summaries. In contrast, existing approaches failed to generate the effective passing test programs that help accurately isolate this faulty file from the others, making the faulty file localized outside the top 20 positions. In our final experiment, we will also evaluate the contribution of the document summarization component to the overall effectiveness of \tool{} in an ablation study.

\subsection{Test Enhancement}
\label{sec:enhance}
Given the large number of source files involved in a test failure, relying solely on the failing test program often fails to provide adequate insights into the cause of the failure. Studies have demonstrated that test coverage can effectively narrow down the search space for faulty files by assigning each file a likelihood of being faulty~\cite{sbfl4}. To enhance this process, \tool{} includes a test enhancement component that aims to generate a set of passing test programs closely related to the failing one, thereby improving the discriminative ability of test programs.

Currently, \tool{} utilizes RecBi~\cite{RecBi} for generating passing test programs, as it has proven effective in producing high-quality test programs that significantly aid in compiler bug isolation through structure mutation operations. Importantly, the test enhancement component in \tool{} is flexible and can be replaced with any established test program generation method, as it operates independently of the core methodology.

 
\subsection{Coverage Analysis}
\label{sec:coverage}
It is well-known that existing coverage-based fault localization methods rely heavily on the quality of test programs. Accurately isolating faulty files is challenging, especially when many source files are involved in failing test executions. For instance, spectrum-based fault localization (SBFL) assigns suspiciousness scores to source files based on their execution across test programs. This task becomes even more difficult with a limited number of test programs, as seen in compiler bug isolation.

For instance, the widely-used SBFL method, Ochiai~\cite{ochiai}, calculates the likelihood of a source file being faulty using the formula presented in Formula~\ref{eq:ochia}. In this formula, \textit{failed(f)} and \textit{passed(f)} represent the numbers of failing and passing test programs covering the source file \textit{f}, while \textit{totalfailed} is the total number of failed test programs. Thus, Ochiai uses this formula to assess the fault probability for each file \textit{f}. Consequently, the effectiveness of fault localization significantly depends on the quantity and quality of the test programs.
 \begin{equation}\label{eq:ochia}
sus(f)=\frac{failed(f)}{\sqrt{totalfailed \times (failed(f) + passed(f))}}
\end{equation}
Consider the GCC bug (ID: 59221) as an example, where only three passing test programs were generated even with the state-of-the-art test enhancement method, RecBi~\cite{RecBi}. Notably, all passing test programs covered the faulty file, resulting in a suspicious score of ${1}/{\sqrt{1\times (1+3)}}=0.5$, ranking it $23^{rd}$ on the list of suspicious files.
In compiler bug isolation, it is common for \textit{totalfailed} and \textit{failed(f)} to be 1, since typically there is only one failing test program per bug. Thus, the performance of traditional SBFL methods is heavily dependent on the number of passing test programs, which remains a challenge as previously mentioned.
Given the limited number of test programs and the multitude of source files involved in the failing execution, exploring new methods to more effectively utilize test programs is crucial. Our analysis reveals that, although passing test programs often cover the same source files as the failing test program, the number of executions varies due to recursive method calls (i.e., one source file could be executed multiple times by one test program). This observation encourages us to explore the potential of using finer-grained execution coverage, as opposed to traditional test coverage used by SBFL, to enhance fault localization in compiler bug isolation scenarios.

\begin{figure}[htbp]
	\centerline{\includegraphics[width=0.88\linewidth]{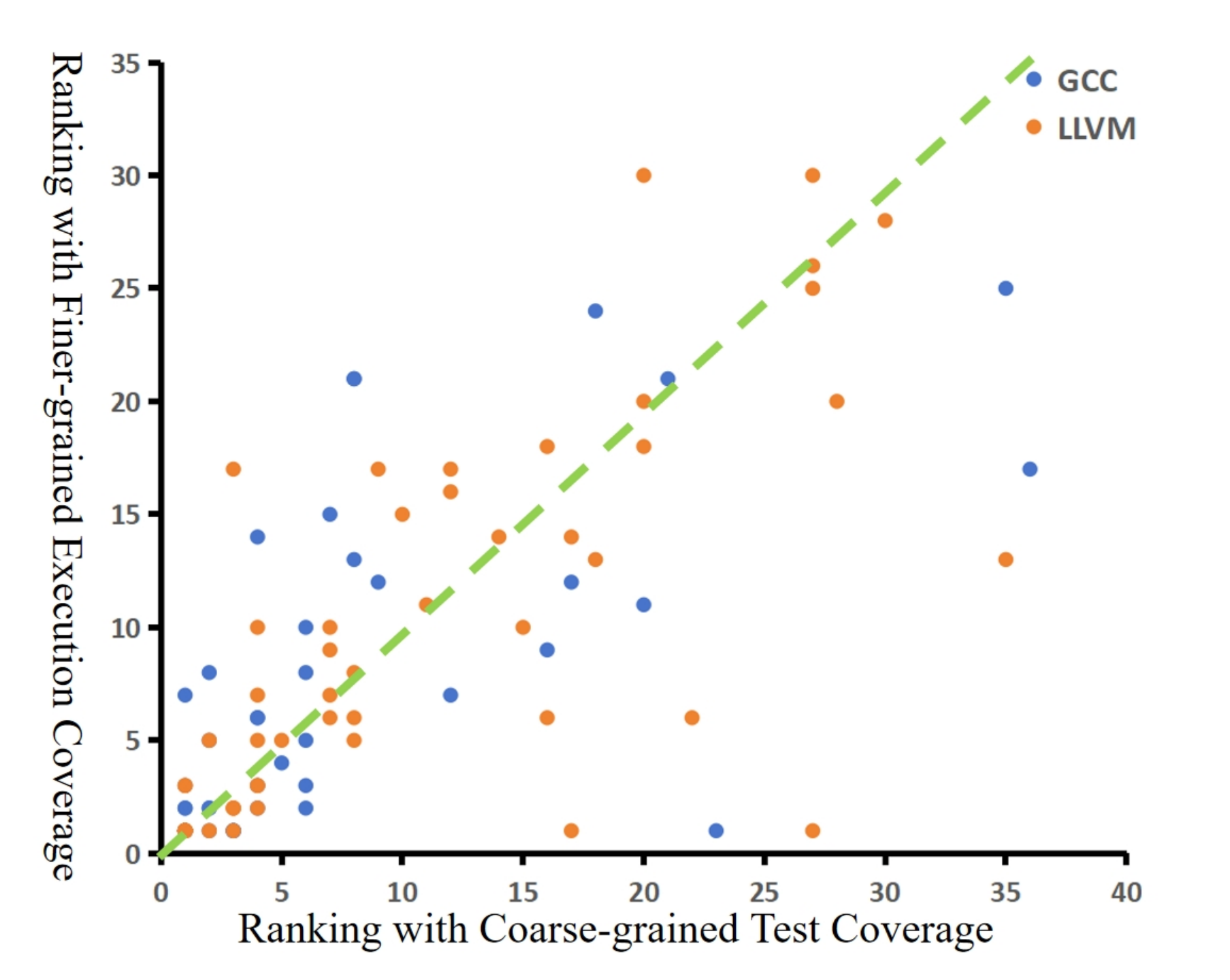}}
	\caption{Fault localization results by Ochiai when using execution coverage against test coverage.}
	\label{fig-coverage}
\end{figure}

\textbf{Preliminary Study:} 
To investigate the fault localization performance using finer-grained execution coverage, we conducted a preliminary study employing the same SBFL formula but incorporating this detailed level of coverage. We selected the Ochiai formula as a representative due to its widespread use in previous studies and proven effectiveness. In this context, \textit{failed(f)} and \textit{passed(f)} denote the number of failing and passing test executions that cover the source file \textit{f}, respectively, while \textit{totalfailed} represents the number of failing test executions involving the file.
Our experiment used a widely-referenced compiler bug dataset~\cite{DiWi,RecBi}, containing 120 compiler bugs from GCC and LLVM. We compared these results with those obtained from traditional SBFL methods. Fig.~\ref{fig-coverage} shows the analysis results, focusing on bugs that could be localized within the top 40 positions for clarity and improved readability. In the figure, the \textit{x-axis} represents the positions of faulty files in the localization results using original coarse-grained test coverage, while the \textit{y-axis} indicates the positions using finer-grained execution coverage. Each point in the figure corresponds to a specific compiler bug from either GCC or LLVM.

The results indicate that finer-grained execution coverage can indeed enhance fault localization accuracy for some compiler bugs-- It improved the localization results for 47/120 compiler bugs compared to coarse-grained test coverage. For instance, in the GCC bug (\href{https://gcc.gnu.org/bugzilla/show_bug.cgi?id=59221}{ID: 59221~\cite{GCCBUGID59221}}), the target faulty file ``{\tt tree-saa-threadupdate.c}'' rose from the $23^{rd}$ position with coarse-grained test coverage to the $1^{st}$ position with finer-grained execution coverage. However, this approach can also reverse results for other bugs. Notably, finer-grained execution coverage worsened fault localization outcomes for 34/120 compiler bugs. For example, the faulty file ``{\tt DAGCombiner.cpp}'' for the LLVM bug (\href{https://bugs.llvm.org/show_bug.cgi?id=25154}{ID: 25154~\cite{LLVMBUGID25154}}) dropped from $1^{st}$ to $10^{th}$ position when using finer-grained coverage.
Hence, both coverage types have their strengths and weaknesses, and they complement each other. Based on this conclusion, \tool{} integrates both types of coverage information to enhance fault localization performance. \tool{} computes suspicious scores for each source file with both coverage types, generating two ranking lists of potential faulty files, which are then fed into LLMs for integration and further refinement together with comprehensive information, such as aforementioned document summaries.

\subsection{Prompt Engineering}
\label{sec:prompt}

To leverage LLMs for integrating comprehensive information in compiler bug isolation, we have crafted a specialized prompt. As motivated in the Introduction, \tool{} currently incorporates four types of information: document summaries of source files involved in the failing test execution, suspicious file lists calculated by both coarse-grained test coverage and finer-grained execution coverage, the failing test program, and the compilation configuration along with related output messages.

The prompt is designed to enable LLMs to understand the file function and assist in re-ranking suspicious files using this comprehensive information, leading to a more precise file ranking. This step is central to the entire \tool{} workflow, leveraging the powerful reasoning and contextual comprehension capabilities of LLMs. By providing detailed prompt information, the model can make nuanced assessments of suspicious files within various contexts.
For example, by analyzing test outputs that trigger the fault under different optimization configurations, the LLM may identify the root cause of the failure and how these optimization strategies impact the program’s execution, further narrowing down the range of potential faulty files. Meanwhile, source file summaries can help LLMs refine the ranking of suspicious files by correlating the root cause with the functions of the source files.

\begin{figure}[htbp]
\centerline{\includegraphics[width=0.89\linewidth]{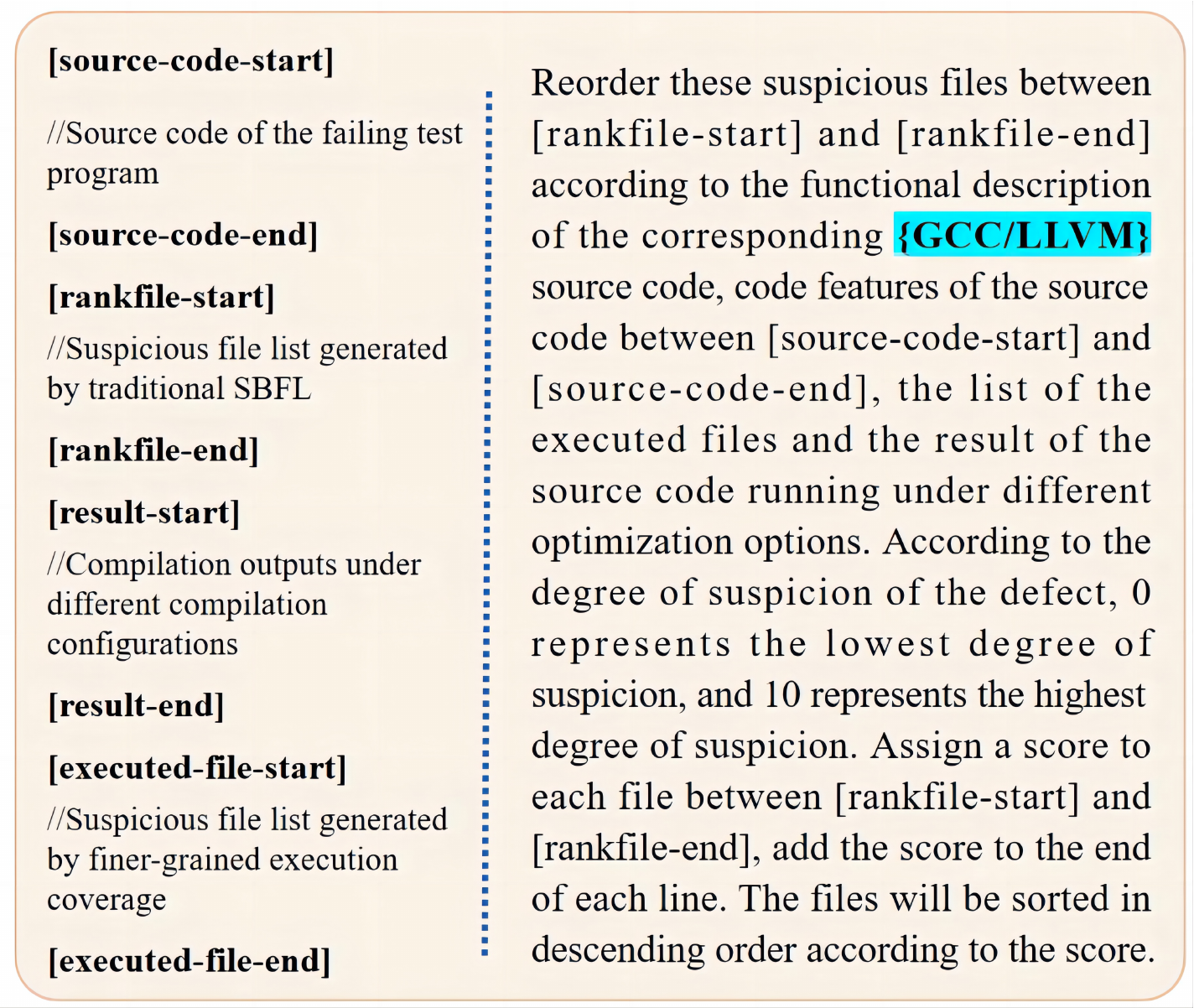}}
\caption{The prompt used by \tool{} for integrating multiple data sources.}
\label{fig6}
\end{figure}
Specifically, we first input the compiler source file summaries into the LLM to ensure the model can understand the specific role of each file in the compiler. This step is crucial because a compiler is a complex system with hundreds of files, each serving unique functions. By providing these function summaries, the model can better understand which files play critical roles in the compilation process, thus more effectively evaluating each file’s potential relevance and suspiciousness in the fault localization task. The prompt for providing document summaries is structured as follows: "\textit{[summary-start]...[summary-end]$\backslash\backslash$ The functional description between [summary-start] and [summary-end] is related to the source code files of \{GCC/LLVM\}. Learn the functions of these files and complete the following tasks based on the functions of these files}". Each file's summary is organized as illustrated in Fig.~\ref{fig5}. 
 Building on this, the detailed prompt is shown in Fig.~\ref{fig6}.
 It is structured into five parts, providing the model with rich contextual information, helping the LLM combine clues from different sources to localize the bug. The specific prompt content is as follows:

\begin{enumerate}[leftmargin=*]
    \item \textbf{Source code of the failing test program}. We first provide the source code of the failing test program that triggers the compilation failure. This information helps the LLM identify the code features in the test program that cause the abnormal behavior or failure during the compilation process, serving as the starting point for subsequent fault localization. In the prompt, we use \textbf{[source-code-start]} and \textbf{[source-code-end]} to mark the beginning and end of the failing test program information.
    \item \textbf{Suspicious file list generated by traditional SBFL}. After enhancing the test programs, \tool{} leverages traditional SBFL techniques to generate a list of suspicious files. This list serves as the target for refinement by the LLM, which synthesizes the insights from various contextual sources to improve localization accuracy. In the prompt, we use \textbf{[rankfile-start]} and \textbf{[rankfile-end]} to mark the beginning and end of the file list.
    \item \textbf{Compilation outputs under different compilation configurations}. Compiler optimization options can significantly influence the output of test programs. To assist in this analysis, we provide the outputs of the failing test program under different configurations. These results enable the LLM to better understand the abnormal behavior observed during compilation and to assess how various optimization strategies impact the code execution path. This, in turn, helps in
    narrowing down the range of potentially faulty files. 
    In the prompt, we use \textbf{[result-start]} and \textbf{[result-end]} to mark the beginning and end of the compilation output under different configurations.
    \item \textbf{Suspicious file list generated by finer-grained execution coverage}. As demonstrated in the preliminary study presented in Section~\ref{sec:coverage}, finer-grained execution coverage information complements the coarse-grained test coverage employed by traditional SBFL methods. Consequently, we also provide the LLM with a list of suspicious files derived from finer-grained coverage data. The expectation is that the LLM will further assess the relationship between these files and the compiler fault based on their execution frequency and coverage. Files with higher coverage may have a higher likelihood of being likely to the bug. In the prompt, we use \textbf{[executed-file-start]} and \textbf{[executed-file-end]} to mark the beginning and end of the file list calculated by utilizing the finer-grained execution coverage.
    
    \item \textbf{Task description}. Finally, we offer a task description that outlines the purpose of each piece of information provided, aiding the LLM in comprehending the components of the prompt and their relevance to the fault localization process. This description includes detailed explanations for each data segment, enabling the model to effectively integrate and analyze data from multiple sources, such as the failing test program, the list of suspicious files, and compilation results. The task description is localized at the end of the prompt as illustrated in Fig.~\ref{fig6}.
    

\end{enumerate}
\section{Evaluation}
\subsection{Research Questions}


\begin{itemize}
    \item \textbf{RQ1}: How effective is \tool{} in compiler bug isolation?
    \item \textbf{RQ2}: What is the impact of different types of information on \tool{}'s performance?
    \item \textbf{RQ3}:
    What is the impact of different SBFL formulas and LLMs used in \tool{} on its performance?
\end{itemize}
 
\subsection{Implementation and Configuration}
\label{sec:implementation}

To assess the performance of our approach, we
developed an automated compiler bug isolation tool, named \tool{}. Specifically, \tool{} integrates the test generation capabilities of the
state-of-the-art RecBi~\cite{RecBi} for test enhancement, which has been shown to be effective by using structural mutations for compiler test program generation. Notably, our approach can be combined with any test generation technique since it functions independently of the test enhancement component.

For the suspicious file list based on coarse-grained test coverage, we adopted the default SBFL formula (i.e., Ochiai) used by RecBi~\cite{RecBi}, as it has performed well in previous studies. In contrast, for the list based on finer-grained execution coverage, we employed the SBFL formula Wong2~\cite{wong2} by default, which achieved slightly better fault localization results. Additionally, we  empirically explore the impact of different formulas on localization results for both coverage types in the final evaluation, and the results show that the formulas have marginal influence to the effectiveness of \tool{}. Table~\ref{SBFL} presents the formulas that will be examined. In particular, each formula has two versions, corresponding to the use of different coverage types: test coverage and execution coverage as illustrated in Table~\ref{SBFL}.

\begin{table}
    \centering
      \caption{SBFL formulas  using different coverage.}
      \resizebox{1\columnwidth}{16mm}{
    \begin{tabular}{c|c|c}
    \toprule
         \multirow{2}{*}{\textbf{Name}}&  \multicolumn{2}{c}{\textbf{Formula}} \\ \cline{2-3}
         & \textbf{w/ Test Coverage} &\textbf{w/ Execution Coverage} \\
    \midrule
    Wong2 & $failed(f)-passsed(f)$ & 
    $C_f(f) - C_p(f) $\\
         Ochiai &  
         $\frac{failed(f)}{\sqrt{totalfailed \times (failed(f) + passed(f))}}$ &
$\frac{ C_f(f)}{ \sqrt{C_f(f) \times ( C_f(f) + C_p(f) ) } }$\\
DStar2& 
$\frac{failed(f)^2}{passed(f)+(totalfailed-failed(f))}$ &$
\frac{ C_f(f)^2}{ C_p(f)+ C_f(f) } $\\
Barinel&  
$1-\frac{passed(f)}{passed(f)+failed(f)}$ &
$1-\frac{ C_p(f)}{ C_p(f) + C_f(f)}$  \\
Tarantula&
$\frac{failed(f)/totalfailed}{failed(f)/totalfailed + passed(f)/totalpassed} $ &
$\frac{ C_f(f)}{ \overline{C}_p(f) + C_f(f) } $\\
\bottomrule
\multicolumn{3}{l}{\small * \textit{C$_f$(f)/C$_p$(f)}: execution times of source file \textit{f} in the failing/passing test runs; }\\
\multicolumn{3}{l}{\small 
* \textit{$\overline{C}_p(f)$}: average execution times among passing test programs that cover source file \textit{f};}\\
    \end{tabular}
  }
    \label{SBFL}
\end{table}

Furthermore, \tool{} utilizes ChatGPT-4o~\cite{ChatGPT-4o} as the default LLM for generating document summaries and making localization inference. This ensures that the model can effectively respond to prompts and perform complex reasoning based on multi-source input data, thereby improving the accuracy of bug localization.




\subsection{Benchmark}

To ensure the comprehensiveness and representativeness of our study, we utilized a commonly-used benchmark dataset containing 120 real-world compiler bugs,
in line with previous research~\cite{RecBi,DiWi,llm4cbi,zhou2022locseq,yang2022isolating}. This dataset includes an equal distribution of 60 GCC~\cite{gcc} bugs and 60 LLVM~\cite{llvm} bugs, covering various types of issues within these compilers. Specifically, each bug linked to the specific compiler version that caused it, along with the failing test programs and compilation configurations that trigger the bug, as well as the actual files responsible for the bug (serving as the ground truth for our compiler faulty file localization task). This data provides a robust foundation for evaluating the effectiveness of \tool{}.

\subsection{Evaluation Metrics}

To evaluate the performance of \tool{} in compiler bug isolation, we employed several commonly used metrics to measure fault localization effectiveness~\cite{RecBi,DiWi,llm4cbi,zhou2022locseq,yang2022isolating}. The specific evaluation metrics are detailed below. Additionally, we will also discuss the \textbf{cost} of using LLMs in our evaluation.

\begin{itemize}[leftmargin=*]
    \item \textbf{Top-N}: This metric measures the number of bugs successfully isolated among the top N positions in the suspicious file list, with N values of 1, 5, 10, and 20. A higher Top-N value indicates better performance, reflecting that more actual faulty files are found within the top N, thus improving bug isolation efficiency.
    \item \textbf{Mean First Rank (MFR)}: 
    This metric reflects the average position of the first identified faulty file in the suspicious file list. A lower MFR indicates quicker identification of the first faulty file, helping developers narrow their debugging scope. Thus, a smaller MFR value signifies better performance by finding faulty files earlier, reducing debugging time and effort.
    
    \item \textbf{Mean Average Rank (MAR)}: 
    When a bug involves multiple faulty files, MAR indicates the average position of all these files in the suspicious file list. For multiple bugs, it averages these values across all cases. A smaller MAR is preferable, as it signifies better accuracy in locating all faulty files, not just the first, and suggests consistent performance when addressing multiple bugs.
    
\end{itemize}
 
\subsection{Baselines}
\label{sec:baseline}

To provide a comprehensive evaluation of our approach's performance, we aimed to compare it with all state-of-the-art compiler bug isolation methods. However, because some existing methods have not open-sourced their implementations and we were unable to replicate their results based on the original papers (including ODFL~\cite{yang2022isolating}, LocSeq~\cite{zhou2022locseq}, and LLM4CBI~\cite{llm4cbi}), we ultimately selected RecBi~\cite{RecBi}, DiWi~\cite{DiWi} and FuseFL~\cite{fusefl} (an LLM-based method) -- three widely used state-of-the-art baselines in previous studies -- as the baselines for our evaluation. Specifically, we perform fault localization at the file level by following existing studies~\cite{yang2022isolating,zhou2022locseq,llm4cbi,RecBi,DiWi}. 
To ensure a fair comparison, we adhered strictly to their experimental settings when running their open-source implementations.  Each tool was executed in five iterations for each bug, selecting the best-performing results to replicate the outcomes reported in their original papers. Particularly, we adopted the same ChatGPT-4o (more powerful than ChatGPT-3.5 originally used by FuseFL) in FuseFL for a fair comparison.



In addition, to evaluate \textbf{the contribution of each component} to the effectiveness of our approach, we also developed a set of variants of \tool{} for comparison. We use \variant{-src} to represent the variant of \tool{} by removing the component of \textit{src} from it. In particular, \textit{src}$\in$ \{\textit{summary, execov, testcov, compile, llm, failtest}\}, which respectively represents the document summary, the suspicious file list derived from finer-grained execution coverage, the suspicious file list derived from coarse-grained test coverage, the compilation results of failing test programs, the large language model (i.e., relying solely on finer-grained execution coverage after removing LLM), and the failing test program.

Furthermore, we assess \textbf{the impact of different formulas} on the effectiveness of \tool{} by creating another two sets of variants that use different formulas for calculating suspicious scores. Specifically, \variant{+FORMULA[exec]} utilizes \textit{FORMULA} for finer-grained execution coverage, while \variant{+FORMULA[test]} applies it for coarse-grained coverage. The formulas considered include Ochiai~\cite{ochiai}, Wong2~\cite{wong2}, DStar2~\cite{Dstar}, Barinel~\cite{barinel}, and Tarantula~\cite{tarantula}, with the default configurations being \variant{+Wong2[exec]} and \variant{+Ochiai[test]}. Finally, to evaluate \textbf{the impact of different LLMs} on the effectiveness of our approach, we developed another two variants of \tool{}, i.e., \variant{+deepseekv3} and \variant{+deepseekr1}, which represent replacing the default LLM ChatGPT-4o in \tool{} with DeepSeek-v3-671B~\cite{DeepSeek-V3} and DeepSeek-R1-671B~\cite{DeepSeek-R1}, respectively.

By employing the above variants for comparison in our experiment, we can comprehensively evaluate the contribution of each component and their influence. This may also provide insights for further refinement in future research.

\section{Results and Analysis}
\subsection{Overall Effectiveness of \tool{}
}

\begin{table*}[tbp]
\caption{Overall effectiveness of \tool{} compared with state-of-the-art baseline methods.}
\begin{center}
\resizebox{\textwidth}{!}{
\begin{tabular}{c|c|cc|cc|cc|cc|cc|cc} \hline 
\toprule
\textbf{Subject}&\textbf{Approach}&\textbf{Top-1}&    \textbf{\textuparrow Top-1}&\textbf{Top-5}&\textbf{\textuparrow Top-5}&\textbf{Top-10}& \textbf{\textuparrow Top-10}& \textbf{Top-20}& \textbf{\textuparrow Top-20}& \textbf{MFR}& \textbf{\textuparrow MFR}& \textbf{MAR}&\textbf{\textuparrow MAR}\\ 
\midrule
\rowcolor{gray!40} \cellcolor{white}
\multirow{4}*{GCC}& \tool{}
& \textbf{20}&     /&\textbf{33}&/& \textbf{39}& /& \textbf{45}& /& \textbf{7.71}& /& \textbf{9.22}&/
\\ 

 &Recbi
& 12&66.67\%&26&26.92\%& 36&
  8.33\%& 41& 9.76\%& 13.69& 43.68\%& 15.62&40.97\%\\ 
 
 &DiWi& 5& 300.00\%& 15& 120.00\%& 26& 50.00\%& 35& 28.57\%& 20.77& 62.88\%& 21.34&56.76\%\\

  &FuseFL
& 10&100.00\%&22&50.00\%& 30&
  30.00\%& 37&21.62\%& 12.57& 38.66\%& 14.49&36.37\%\\ 
 \midrule
\rowcolor{gray!40} \cellcolor{white}
\multirow{4}*{LLVM}& \tool{}
& \textbf{22}& /& \textbf{30}& /& \textbf{39}& /& \textbf{50}& /& \textbf{8.70}& /& \textbf{10.25}&/
\\
 & 
Recbi
& 13& 69.23\%& 26& 15.38\%& 35& 11.43\%& 48& 4.17\%& 10.58& 17.77\%& 12.1&15.29\%\\
 & DiWi& 5& 340.00\%& 19& 57.89\%& 28& 39.29\%& 38& 31.58\%& 17.98& 51.61\%& 18.84&45.59\%\\

   &FuseFL
& 14&57.14\%&26&15.38\%& 33&
  18.18\%& 41&21.95\%& 13.31& 34.64\%& 13.43&23.68\%\\ 
\bottomrule
\multicolumn{14}{l}{*Columns "\textuparrow" present the improvement rates of \tool{} over a compared technique in terms of various measurements.}\\
\end{tabular}
}
\label{tab1}
\end{center}
\end{table*}

Table~\ref{tab1} presents the experimental results of our approach compared to baseline methods on the benchmark dataset. The findings demonstrate that \tool{} significantly outperforms state-of-the-art baselines across all metrics.
Specifically, \tool{} successfully localized 20, 33, 39, and 45 GCC compiler bugs within the Top-1, Top-5, Top-10, and Top-20 positions, respectively, on the suspicious file list refined by the LLM after integrating multiple data sources. It improves over RecBi by 66.67\%, 26.92\%, 8.33\%, and 9.76\% in these rankings, indicating \tool{}’s effectiveness in narrowing down the scope of faulty files, particularly excelling in the Top-1 and Top-5 positions. Notably, Top-1 and Top-5 are crucial metrics for localization techniques, as developers typically prefer to examine no more than five suggested locations provided by automated methods, as highlighted in previous studies~\cite{practitioners2016Kochhar}.
Similarly, \tool{} localized 22, 30, 39, and 50 LLVM compiler bugs within the Top-1, Top-5, Top-10, and Top-20 positions, achieving improvements of 69.23\%, 15.38\%, 11.43\%, and 4.17\% over RecBi. Compared to DiWi and FuseFL, the improvements in Top-1 accuracy range from 300\% to 340\% and from 57.14\% to 100\%.
These results further demonstrate \tool{}’s effectiveness and highlight the generality of our approach for localizing bugs across various compilers.

\begin{figure}[tbp]
\centering
\begin{subfigure}{0.15\textwidth}
    \includegraphics[width=\textwidth]{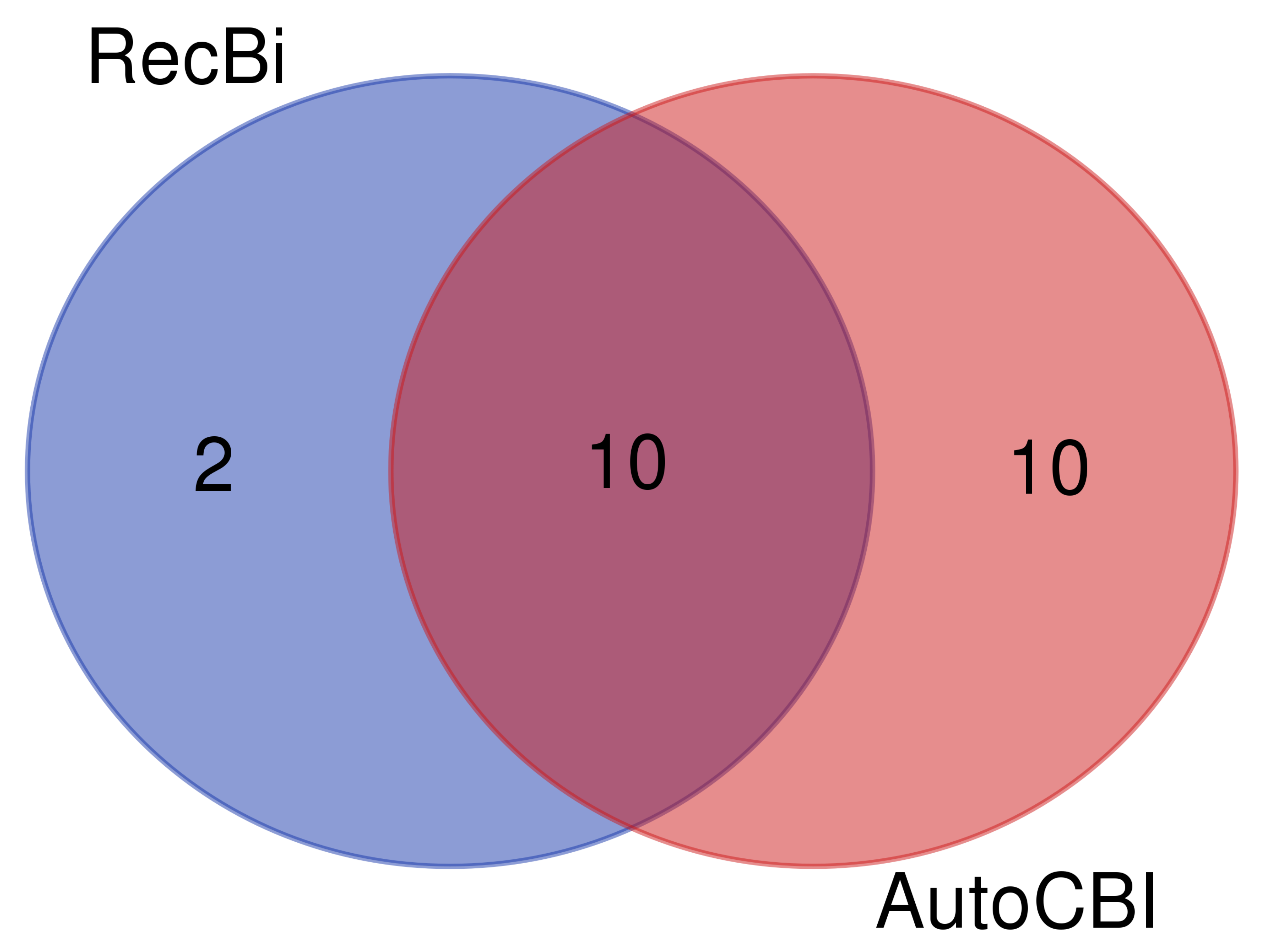}
    \caption{GCC bugs}
    \label{fig-gcc-diwi}
\end{subfigure}
\begin{subfigure}{0.15\textwidth}
\includegraphics[width=\textwidth]{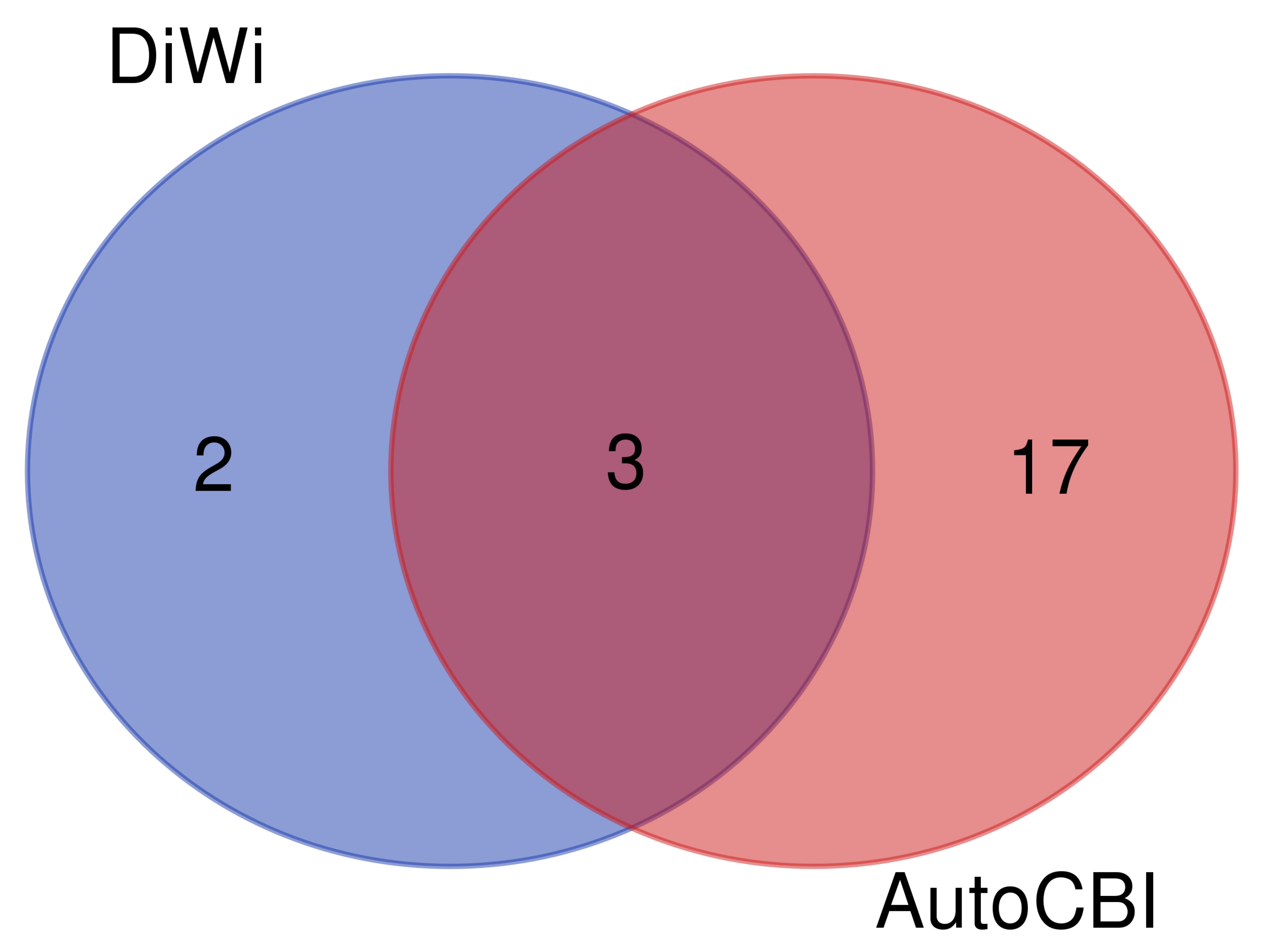}
    \caption{GCC bugs}
    \label{fig-GCC-diwi}        
\end{subfigure}
\begin{subfigure}{0.15\textwidth}
\includegraphics[width=\textwidth]{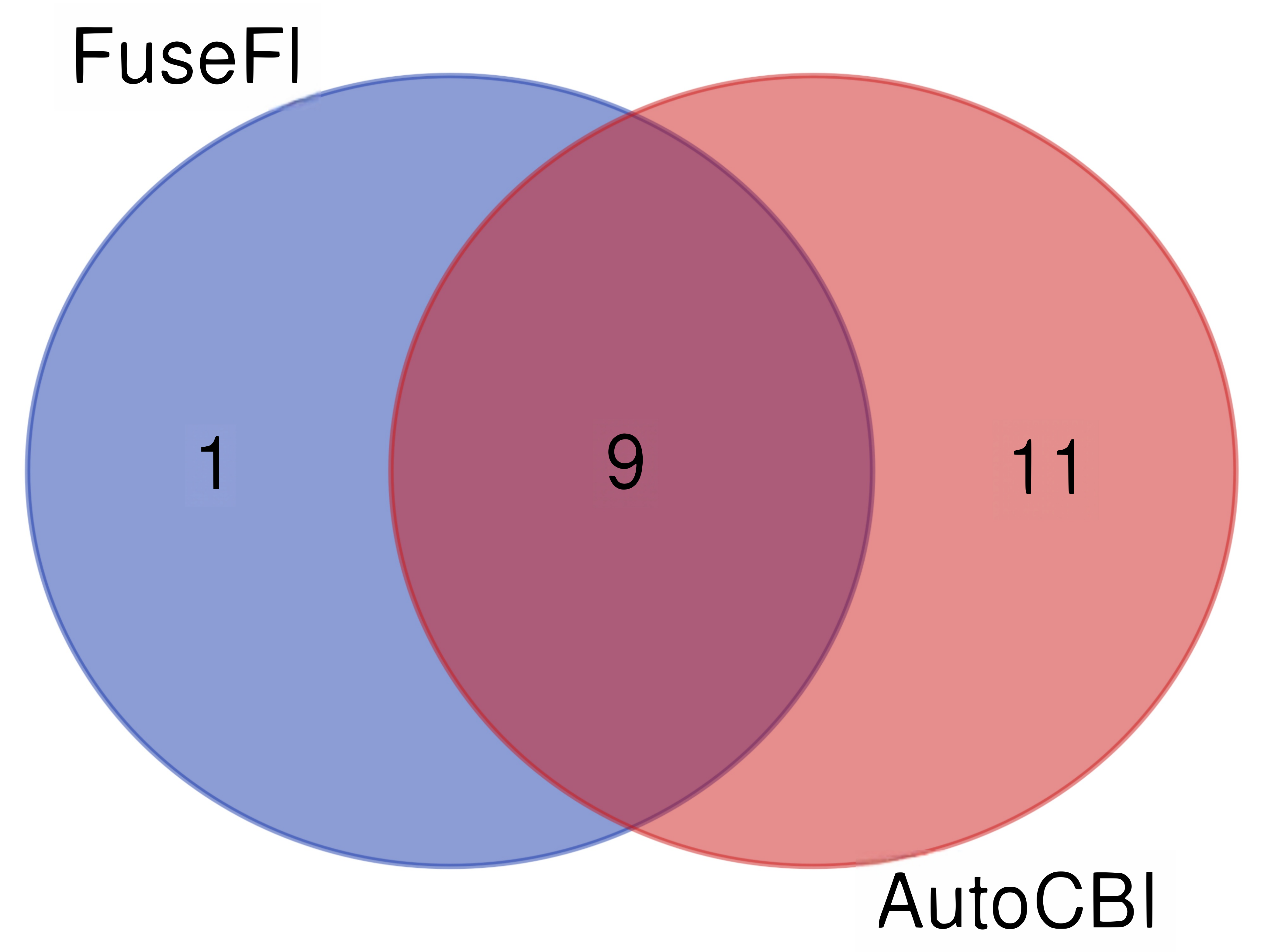}
    \caption{GCC bugs}
    \label{fig-GCC-fusefl}        
\end{subfigure}
\begin{subfigure}{0.15\textwidth}
\includegraphics[width=\textwidth]{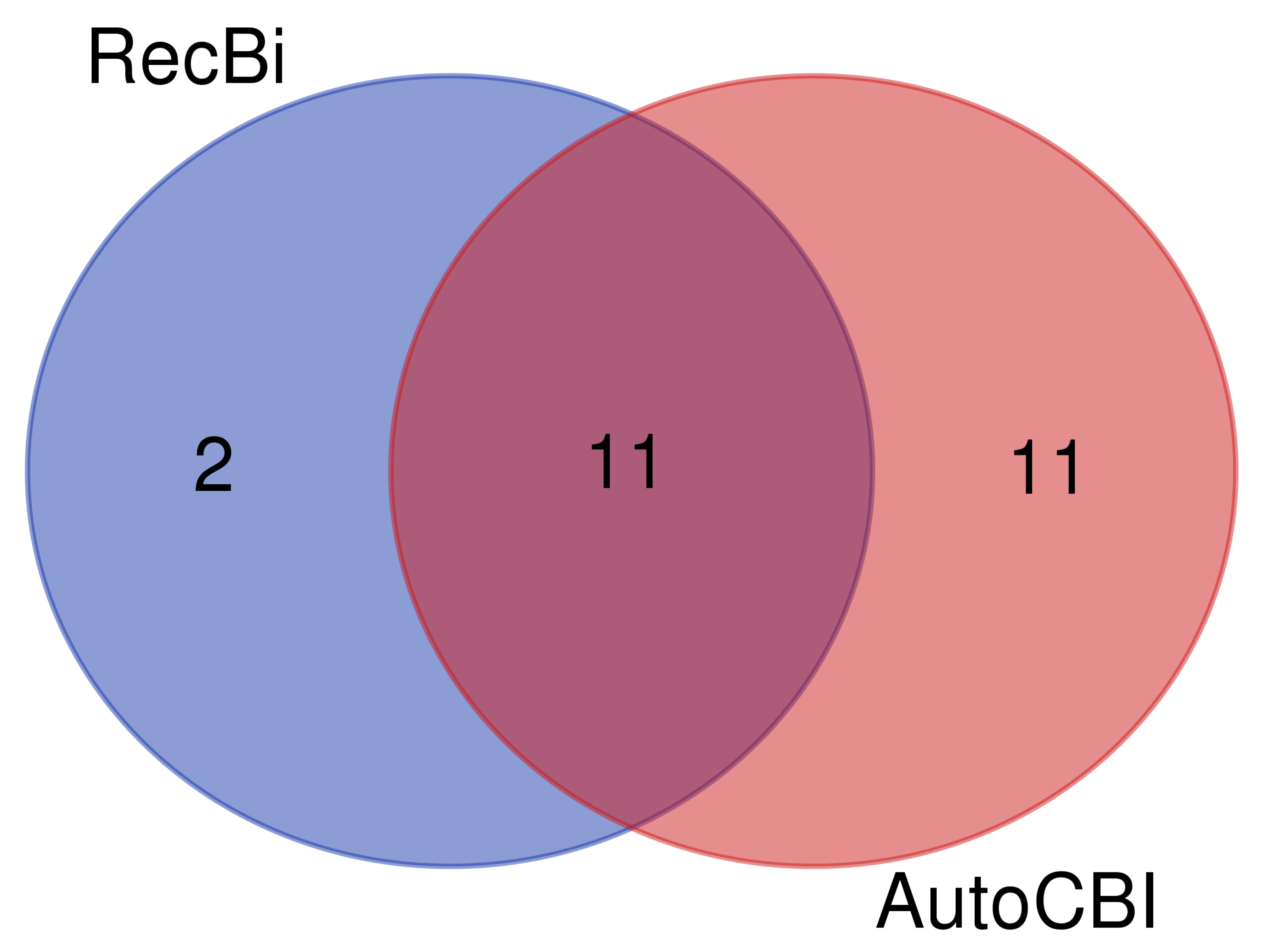}
    \caption{LLVM bugs}
    \label{fig-LLVM-recbi}        
\end{subfigure}
\begin{subfigure}{0.15\textwidth}
\includegraphics[width=\textwidth]{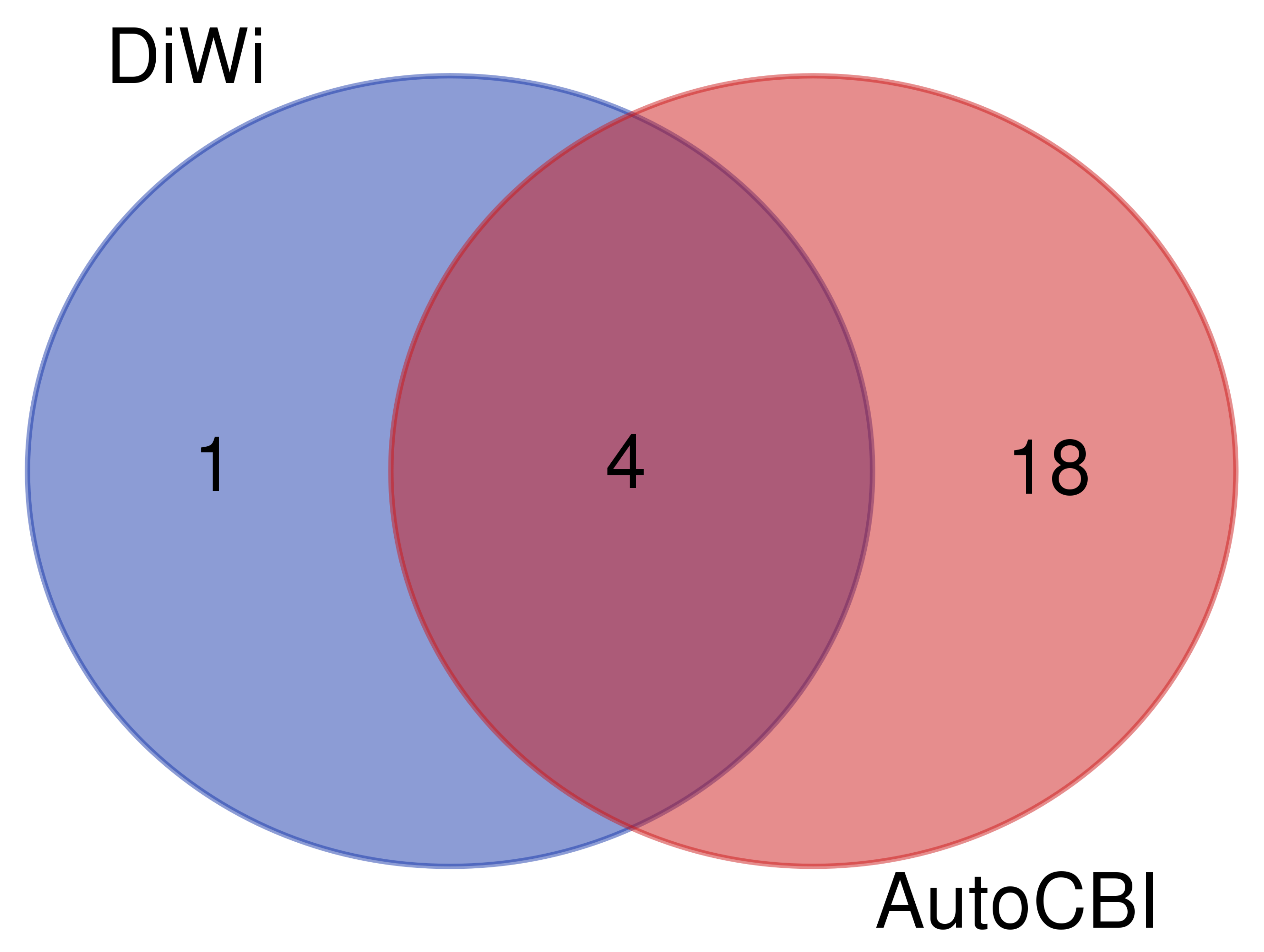}
    \caption{LLVM bugs}
    \label{fig-llvm-diwi}        
\end{subfigure}
\begin{subfigure}{0.15\textwidth}
\includegraphics[width=\textwidth]{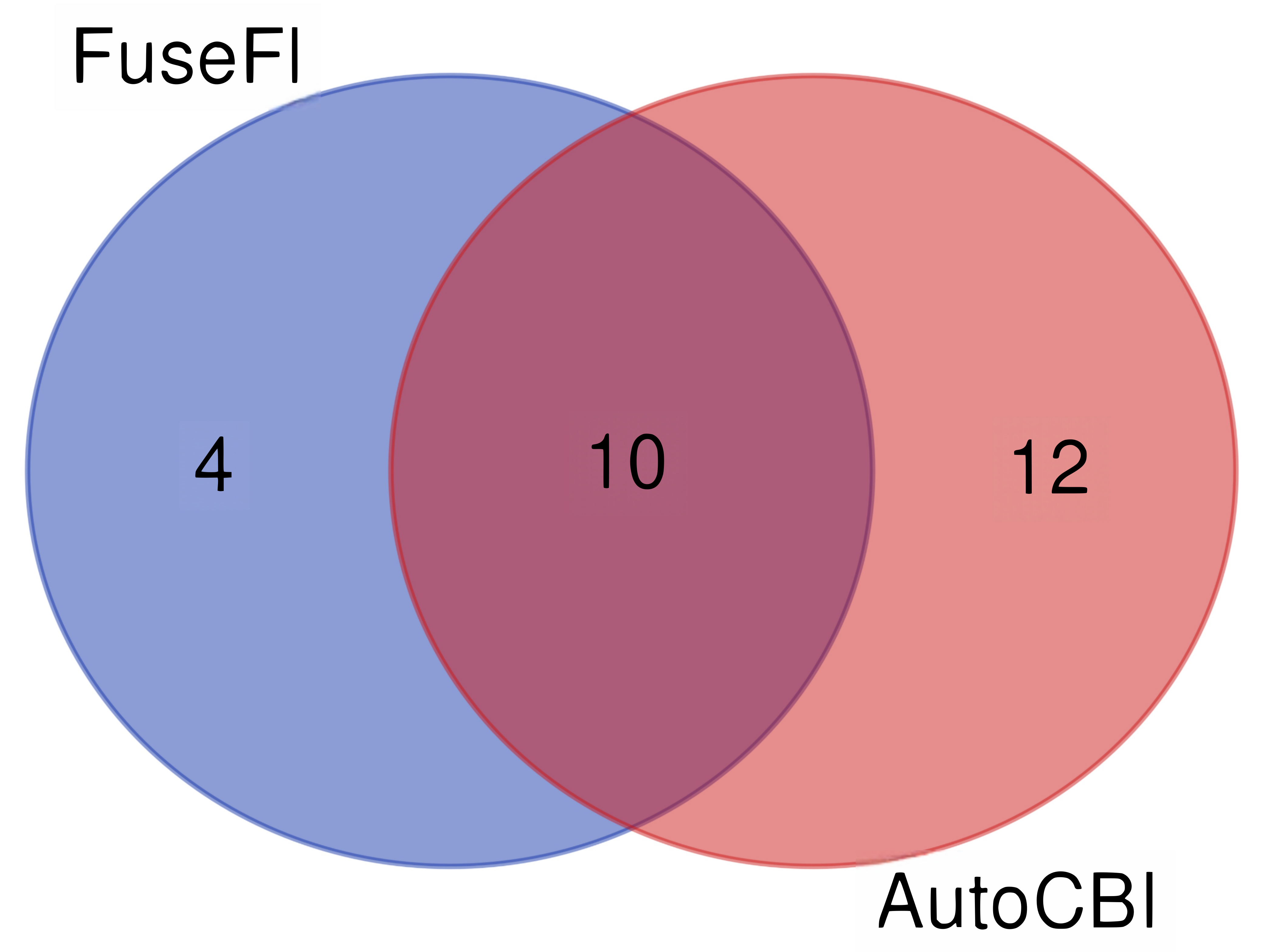}
    \caption{LLVM bugs}
    \label{fig-llvm-fusefl}        
\end{subfigure}
\caption{Overlaps of compiler bugs localized at Top-1.}
\label{fig-methods}
\end{figure}

We also conducted a detailed analysis of the bugs that each method accurately localized within the Top-1 position and examined the overlaps among them. 
Fig.~\ref{fig-methods} illustrates the results. From the figure, we can observe that a substantial portion of the bugs localized by RecBi, DiWi and FuseFL can also be accurately localized by our approach, highlighting the remarkable performance of \tool{}. One key reason for this is that \tool{} incorporates the strengths of RecBi’s test program generation capabilities, which is a core aspect of RecBi. Furthermore, the results demonstrate that integrating multiple data sources significantly enhances the accuracy of compiler bug localization without largely diminishing the effectiveness of individual data sources. 

The significant improvement can be attributed to \tool{}’s ability to integrate multiple data sources and utilize the robust reasoning and understanding capabilities of the LLM. This enables a more efficient analysis of the differences between failing and passing test executions, as well as the incorporation of contextual information like source file summaries. These additional inputs help the LLM more accurately assess the suspiciousness of each file, allowing for more accurate identification and localization of files most likely responsible for compiler bugs. 
For instance, as demonstrated in Section~\ref{sec:motivate}, the GCC compiler bug (\href{https://gcc.gnu.org/bugzilla/show_bug.cgi?id=59221}{ID: 59221~\cite{GCCBUGID59221}}) was initially localized at the 23$^{rd}$ position by the state-of-the-art RecBi. However, by utilizing finer-grained execution coverage, the faulty file was successfully localized at the 1$^{st}$ position.
This allows the LLM to refine the results by considering all relevant factors.


Moreover, we also evaluated the efficiency and cost of our approach. On average, the bug isolation process takes approximately \textbf{275 seconds} per bug, excluding the test enhancement process, and the cost for bug isolation is about \textbf{\$ 0.075} per bug for using ChatGPT-4o. This overhead is relatively minor compared to the test generation and iterative testing processes employed by existing compiler bug isolation methods, and the cost is also relatively small. In summary, \tool{}’s performance demonstrates that integrating multiple data sources with the reasoning and understanding capabilities of large language models can substantially enhance the efficiency of compiler bug isolation. It not only excels in the GCC compiler but also achieves outstanding results in the LLVM compiler. The more precise localization results can help significantly reduce debugging time, providing developers with a more efficient tool that advances intelligent debugging and maintenance of compilers.





\subsection{The Contribution of Each Component}
\label{sec:rq2}
As mentioned in Section~\ref{sec:baseline}, to evaluate the contribution of each component in our approach, 
we have developed a set of variants of \tool{}, and then we performed the same experiment with them and compared their results with the original \tool{}.
Table~\ref{tab2} presents these experimental results (We only report Top-1/5 to save space. The complete results can be found at \href{https://anonymous.4open.science/r/AutoCBI-D823/data/complete%20data.pdf}{our project homepage}), along with the relative improvement of each variant compared to \tool{} (negative numbers indicate performance degradation). The results indicate that all components of \tool{} significantly enhance our approach’s overall effectiveness. Specifically, removing one of the components in \tool{} (i.e., source file summaries, compilation configurations and results, finer-grained execution coverage, coarse-grained test coverage, failing test program and the use of LLM) results in an average decrease of 50.00\%, 45.24\%, 40.48\%, 30.95\%, 54.76\% and 28.57\% in the number of bugs localized within the Top-1 position across the two compilers.


\begin{table}[tbp]
\caption{\tool{}'s result when disabling each component.}
\begin{center}
\resizebox{\columnwidth}{!}{
\setlength\tabcolsep{3pt}
\begin{tabular}{c|c|cc|cc|cc|cc} 

\toprule
\textbf{Sub.}&\textbf{Approach}&\textbf{Top-1}&\textbf{\textuparrow Top-1}&     \textbf{Top-5}&\textbf{\textuparrow Top-5}&\textbf{MFR}&\textbf{\textuparrow MFR}&\textbf{MAR}&\textbf{\textuparrow MAR}\\ 
\midrule
\rowcolor{gray!40} \cellcolor{white}
\multirow{7}*{GCC}& \tool{}
& \textbf{20} &/&     \textbf{33} &/&\textbf{7.71} &/&\textbf{9.22} &/\\ 

& \variant{-summary}& 10 &-50.00\%&      25 &-24.24\%&12.60  &-63.42\%&16.62
 &-80.26\%\\ 
& \variant{-compile}& 9 &-55.00\%& 26 &-21.21\%&13.41  &-73.93\%&
 17.2
 &-86.55\%\\
 & \variant{-execov}& 11 &-45.00\%& 21 &-36.36\%& 11.81  &-53.18\%&15.64
 &-69.63\%\\
  & \variant{-testcov}& 14 &-30.00\%&      25 &-24.24\%&10.06  &-30.48\%&14.28
 &-54.88\%\\ 
& \variant{-llm}& 14&-30.00\%& 23&-30.30\%& 12.08&-56.68\%&16.06&-74.19\%\\
& \variant{-failtest}& 9 &-55.00\%&      21 &-36.36\%&12.56  &-62.91\%&16.60
 &-80.04\%\\
\midrule
\rowcolor{gray!40} \cellcolor{white}
 \multirow{7}*{LLVM}& \tool{}
& \textbf{22} &/& \textbf{30} &/& \textbf{8.70} &/&\textbf{10.25} 
 &/\\
 & 
\variant{-summary}
& 11 &-50.00\%& 21 &-30.00\%& 12.57  &-44.48\%&12.65 
 &-23.41\%\\
 & \variant{-compile}
& 14 &-36.36\%& 24 &-20.00\%& 10.82  &-24.37\%&11.53 
 &-12.49\%\\
 & \variant{-execov}
& 14 &-36.36\%& 24 &-20.00\%& 12.60  &-44.83\%&12.79 
 &-24.78\%\\
  & \variant{-testcov}& 15 &-31.82\%&      26 &-13.33\%&9.35  &-7.47\%&10.74
 &-4.78\%\\ 
& \variant{-llm}& 16 &-27.27\%& 25 &-16.67\%& 9.73&-11.84\%&11.12&-8.47\%\\
& \variant{-failtest}& 10&-54.55\%&      21 &-30.00\%&12.80  &-47.13\%&13.05
 &-27.32\%\\ 
 \bottomrule
\multicolumn{10}{l}{*Columns "\textuparrow" present the improvement rates of experiment variants over \tool{}.}\\
\end{tabular}
}
\label{tab2}
\end{center}
\end{table}

\begin{figure}[tbp]
\centering
\begin{subfigure}{0.242\textwidth}
    \includegraphics[width=\textwidth]{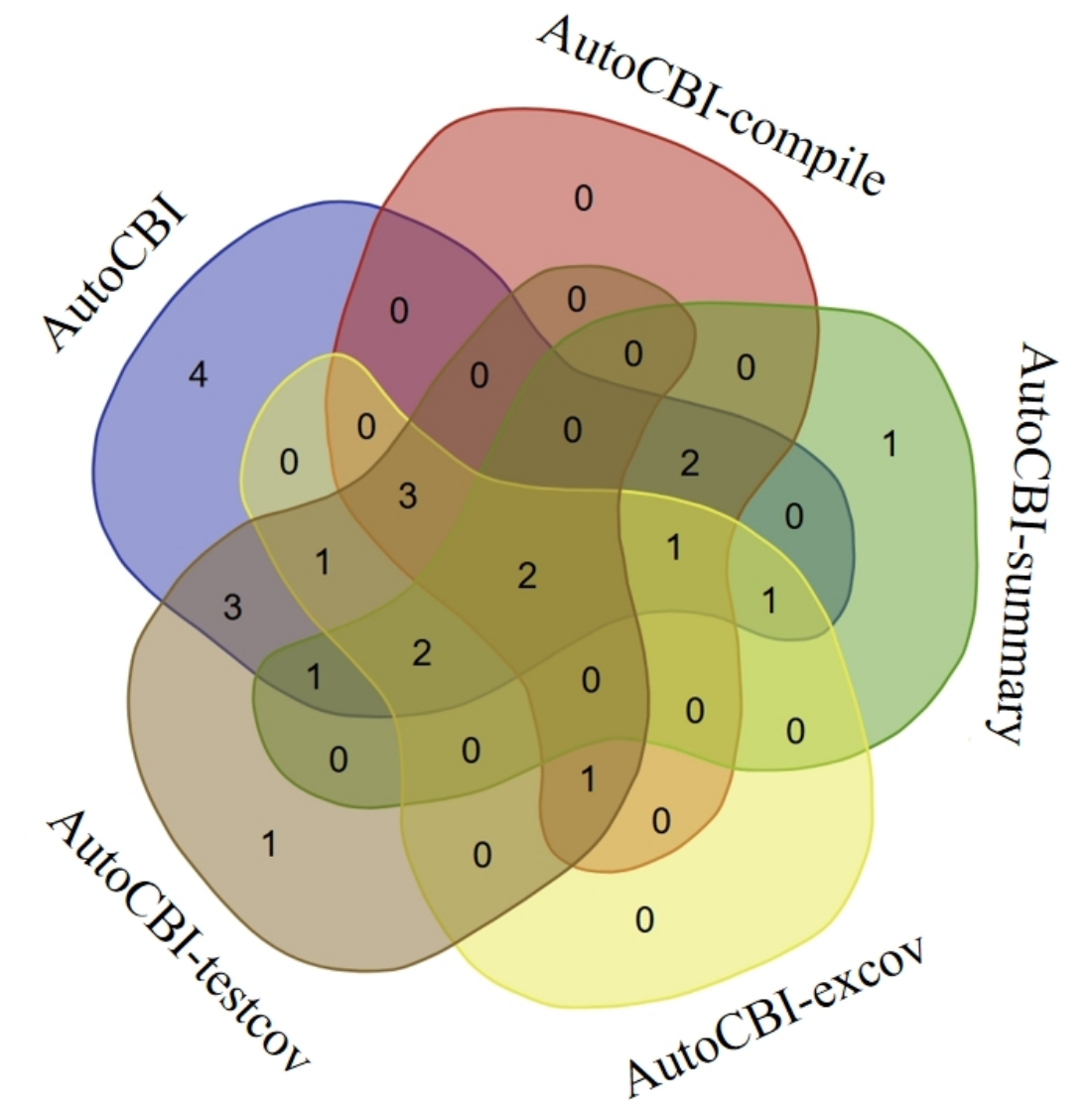}
    \caption{GCC bugs}
    \label{fig-gcc}
\end{subfigure}
\begin{subfigure}{0.229\textwidth}
\includegraphics[width=\textwidth]{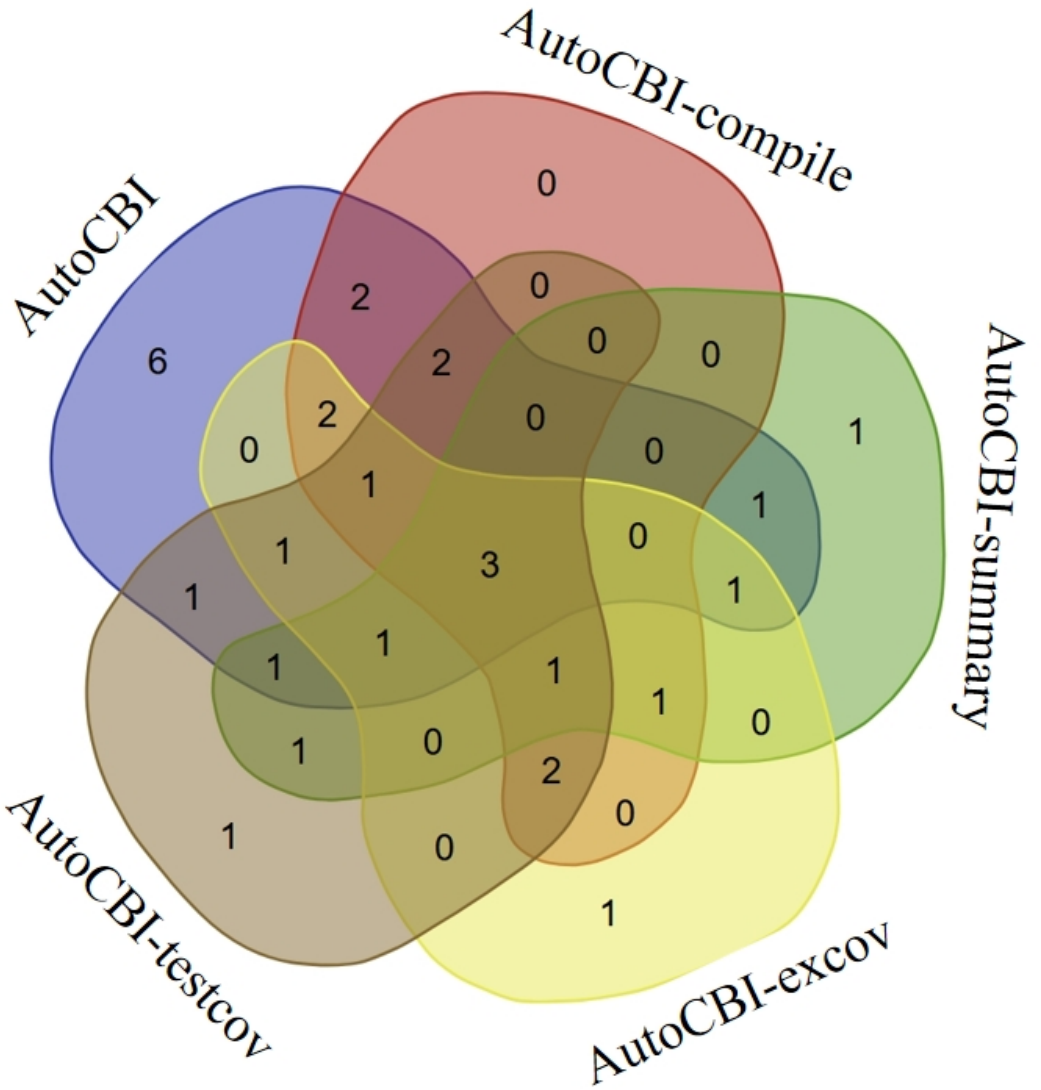}
    \caption{LLVM bugs}
    \label{fig-llvm}        
\end{subfigure}
\caption{Overlaps of compiler bugs localized at Top-1 when using different variants of \tool{}.}
\label{fig-variant}
\end{figure}


Specifically, without the inclusion of source file summaries or the failing test program, \variant{-summary} and \variant{-failtest} are able to localize only half (\textit{vs} \tool{}) of the compiler bugs within the Top-1 position, making the most significant decline in performance of \tool{} compared with the other variants. This highlights that function summaries of source files are essential for the LLM to comprehend the role and the function of each file within the compiler architecture. These summaries provide the LLM with critical background knowledge, enabling the model to not only rely on the surface-level coverage information but also assess a file’s importance and relevance from a functional perspective during the compilation process. This finding underscores that compiler file summaries have a pivotal impact on bug localization and can greatly assist the LLM in better identifying and isolating faulty files. 
Similarly, the failing test program offers the bug-triggering features, which are closely associated with a certain part of compiler implementations and thus can aid the LLM to effectively identify the most possibly faulty locations. Therefore, they should not be overlooked in the compiler bug localization process.

Additionally, the table reveals that when the use of LLM is removed and only finer-grained execution coverage is employed for fault localization, \variant{-llm} successfully localized 14 and 16 bugs within the Top-1 position across the two compilers, still outperforming the baseline methods. This underscores the importance of incorporating finer-grained coverage, particularly when working with a limited number of test programs for compiler bug isolation.

Finally, to investigate the complementariness of the data sources used in \tool{}, we analyzed the overlaps of bugs that each source successfully localized within the Top-1 position. Fig.~\ref{fig-variant} depicts the findings, showing that different data sources complement one another, each contributing to the accurate localization of various compiler bugs. However, an intriguing phenomenon was also observed: adding certain data sources can potentially diminish localization performance for some bugs. For instance, removing finer-grained execution coverage led to five additional bugs being localized within the Top-1 position on the benchmark. This is consistent with our preliminary study in Section~\ref{sec:coverage}, where finer-grained execution coverage occasionally reduced fault localization accuracy. Taking the LLVM compiler bug (\href{https://bugs.llvm.org/show_bug.cgi?id=25154}{ID: 25154~\cite{LLVMBUGID25154}}) as an example, the faulty file ``DAGCombiner.cpp'' was initially ranked 1$^{st}$, but its rank dropped to 10$^{th}$ when using finer-grained execution coverage, which misled the LLM and hindered accurate localization. Similarly, other data sources might also provide inaccurate information affecting \tool{}’s overall performance. For example, removing file summaries allowed five new bugs to be localized within the Top-1 position, likely because the summaries were not informative enough or contained ambiguous descriptions that misled the LLM in correlating failures with source files.

In summary, removing one or more data sources contributed to nine additional bugs being localized within the Top-1 position, as shown in Fig.~\ref{fig-variant}. This suggests that while multi-source information generally enhances model performance, the quality, accuracy, and completeness of this information are crucial for effective localization. In some cases, incomplete or biased information may lead to incorrect localization, and reducing such information can sometimes improve fault localization performance. This indicates the potential for developing an adaptive integration method that automatically determines the usefulness of each data source for fault localization. We plan to explore this approach in future work.

\subsection{Performance under Different Configurations}

\begin{table}[tbp]
\begin{center}
\caption{\tool{}'s result under different configurations.}
\resizebox{\columnwidth}{!}{
\setlength\tabcolsep{3pt}
\begin{tabular}{c|c|cc|cc|cc|cc} 

\toprule
\textbf{Sub.}&\textbf{Approach}&\textbf{Top-1}&\textbf{\textuparrow Top-1}&     \textbf{Top-5}&\textbf{\textuparrow Top-5}&\textbf{MFR}&\textbf{\textuparrow MFR}&\textbf{MAR}&\textbf{\textuparrow MAR}\\ 
\midrule
\rowcolor{gray!40} \cellcolor{white}
\multirow{11}*{GCC}& \tool{}
& \textbf{20} &/&     33 &/&\textbf{7.71} &/&\textbf{9.22} &/\\ 

& \variant{+Ochiai[exec]}& 18 &-10.00\%&      30 &-9.09\%&10.21  &-32.43\%&14.28
 &-54.88\%\\ 

 & \variant{+Barinel[exec]} & 18 &-10.00\%& 28 &-15.15\%&9.75  &-26.46\%&
 11.76
 &-27.55\%\\
 & \variant{+DStar2[exec]}& 17 &-15.00\%& 27 &-18.18\%& 10.60  &-37.48\%&12.54
 &-36.01\%\\
 & \variant{+Tarantula[exec]}& 17 &-15.00\%& 28 &-15.15\%& 10.58  &-37.22\%&14.76
 &-60.09\%\\
\cline{2-10}
& \variant{+Wong2[test]}
& 19
&-5.00\%&      32
&-3.03\%&8.81
&-14.27\%&9.28
&-0.56\%\\ 
 & \variant{+Barinel[test]}& 19
&-5.00\%& 32
&-3.03\%&8.22
&-6.61\%&
 9.36
&-1.52\%\\
 & \variant{+DStar2[test]}& \textbf{20}&\textbf{0.00\%}& 32
&-3.03\%& 8.33
&-8.04\%&9.72
&-5.42\%\\
 & \variant{+Tarantula[test]}& 18
&-10.00\%& 33&0.00\%& 8.08
&-4.80\%&9.64
&-4.56\%\\ \cline{2-10}

  & \variant{+deepseekv3}& 16 &-20.00\%&     34 &3.03\%&10.76  &-39.56\%&12.13
 &-31.56\%\\ 
  & \variant{+deepseekr1}& 16 &-20.00\%&     \textbf{35} &\textbf{6.06\%}&9.96  &-29.18\%&11.5
 &-24.73\%\\ 
 
\midrule
\rowcolor{gray!40} \cellcolor{white}
 \multirow{11}*{LLVM}& \tool{}
& \textbf{22} &/& 30 &/& 8.70 &/&10.25 
 &/\\
 & 
\variant{+Ochiai[exec]}
& 15 &-31.82\%& 27 &-10.00\%& 9.04  &-3.91\%&9.88 
 &3.61\%\\
 & \variant{+Barinel[exec]} 
& 18 &-18.18\%& 30&0.00\%& \textbf{8.24} &\textbf{5.29\%}&\textbf{8.29} &\textbf{19.12\%}\\
 & \variant{+DStar2[exec]}
& 19 &-13.64\%& 29 &-3.33\%& 9.15  &-5.17\%&9.16 
 &10.63\%\\
 & \variant{+Tarantula[exec]}& 18 &-18.18\%& 27 &-10.00\%& 9.63  &-10.69\%&11.17 
 &-8.98\%\\
\cline{2-10}
& 
\variant{+Wong2[test]}
& 21
&-4.55\%& 30&0.00\%& 8.72 
&-0.23\%&10.28 
&-0.29\%\\
 & \variant{+Barinel[test]}& \textbf{22}&\textbf{0.00\%}& 30&0.00\%& 8.46&2.76\%&10.02&2.24\%\\
 & \variant{+DStar2[test]}
& 21
&-4.55\%&30&0.00\%& 8.74 
&-0.46\%&10.30 
&-0.49\%\\
 & \variant{+Tarantula[test]}& 21
&-4.55\%& 29
&-3.33\%& 8.63 
&0.80\%&10.20 
&0.49\%\\ \cline{2-10}

   & \variant{+deepseekv3}& 16 &-27.27\%&     28 &-6.67\%&10.08  &-15.86\%&11.62
 &-13.37\%\\ 
  & \variant{+deepseekr1}& 16 &-27.27\%&    \textbf{31} &\textbf{3.33\%}&9.67  &-11.15\%&11.05
 &-7.80\%\\
\bottomrule
\multicolumn{10}{l}{*Columns "\textuparrow" present the improvement rates of experiment variants over \tool{}.}\\
\end{tabular}
}
\label{tab3}
\end{center}
\end{table}

As discussed in Section~\ref{sec:implementation}, the current implementation of \tool{} uses the Ochiai and Wong2 formulas as default configurations to calculate suspicious file lists based on coarse-grained test coverage and finer-grained execution coverage, respectively. To examine the impact of these configurations, we developed two sets of \tool{} variants, as outlined in Section~\ref{sec:baseline}, where each formula was replaced with an alternative one. Table~\ref{SBFL} provides details on all the formulas studied in this experiment.

The experimental results are presented in Table~\ref{tab3}, where we show the outcomes for each variant and its relative improvement over the default \tool{} (equivalent to \variant{+Wong2[exec]} and \variant{+Ochiai[test]}). Negative numbers indicate a degradation in fault localization performance. The results demonstrate that finer-grained execution coverage is more sensitive to the choice of formula than coarse-grained test coverage. Specifically, changing the formula for test coverage results in an average performance decline of 4.2\% (i.e., one fewer bug localized) in terms of Top-1 accuracy, compared to a decline of up to 16.48\% (i.e., about four fewer bugs localized) when altering the formula for execution coverage. This is because the limited number of test programs and their similar execution paths tend to have a smaller impact on calculation results across different formulas. In contrast, the finer-grained execution coverage of different source files can vary more significantly than test coverage, making it more sensitive to coverage utilization methods. Despite this, the overall performance of \tool{} with different formulas is not significantly affected, and it consistently outperforms baseline methods by a large margin.

We also analyzed the bugs successfully localized by \tool{} under different configurations. The findings support a similar conclusion: different configurations (or formulas) complement each other, as each formula contributes to uniquely localizing different bugs in the Top-1 position. For instance, with finer-grained execution coverage, the Wong2 formula successfully localized five unique bugs that other formulas could not. Similarly, the Tarantula formula contributed to the unique localization of four additional bugs. This suggests that combining the strengths of various formulas, or developing more effective formulas tailored for finer-grained execution coverage (such as the formulas in Table~\ref{SBFL}), could further enhance the fault localization performance of \tool{}.

Finally, to evaluate the performance of \tool{} when utilizing different LLMs, we replaced the default ChatGPT-4o model with DeepSeek-V3 and DeepSeek-R1 as explained in Section~\ref{sec:baseline}. The results are presented in Table~\ref{tab3}. The results demonstrate that their Top-5 performance is similar to ChatGPT-4o, while the Top-1 performance slightly decreased. Nevertheless, our approach always outperforms the baselines. In particular, the cost is much lower, with DeepSeek-V3 requiring only \$0.0045 per bug on average, and DeepSeek-R1 costing \$0.009 per bug.

\section{Discussion}
\subsection{Threats to Validity}

The threats to internal validity mainly arise from the implementation details of \tool{} and its comparative methods -- DiWi~\cite{DiWi}, RecBi~\cite{RecBi}, and FuseFL~\cite{fusefl}. To address this threat, we thoroughly reviewed our implementation of \tool{}. For the baseline methods, we directly used their open-source implementations and successfully replicated the results reported in their original papers. Additionally, we have open-sourced all our implementations to allow for further verification.

The threats to external validity are primarily related to the subjects and the LLMs used in our experiment. For the subjects, we adopted a widely-used benchmark dataset comprising 120 real-world bugs from two mature compilers, GCC and LLVM. While these bugs are diverse and representative, our approach may perform differently on entirely new datasets. 
Regarding the LLMs, We employed ChatGPT-4o as our primary LLM becuase of its strong inference capabilities. 
We also explored two other LLMs DeepSeek-V3 and DeepSeek-R1. The results indicate that while effectiveness varied slightly among LLMs, our approach consistently outperformed all baseline methods. 
Additionally, the potential for data leakage from LLM training involving similar compiler bugs could affect model performance and the fairness of our experiment. However, the distinct focus of our work compared to the training process helps mitigate this threat. Nevertheless, we plan to compile a new dataset of recent compiler bugs to further evaluate our approach's performance.



\subsection{Future Work}

Although our approach yielded promising results in the evaluation, we have identified several exciting directions that can be explored for further improving the performance of \tool{}. 

\textbf{Interactive Bug Isolation}: As discussed in Section~\ref{sec:rq2}, while multiple data sources enhance the overall effectiveness of \tool{}, some may adversely impact fault localization for certain bugs. By interactively incorporating different data sources and using feedback to refine the combination strategy through methods like CoT (Chain-of-Thought) or ToT (Tree-of-Thought), \tool{} can better leverage contextual information from different data sources, leading to further improvement.


\textbf{Non-Prompt-Based Bug Isolation}: 
The prompt provides the LLM with essential information, such as task description and context. An inaccurate prompt can lead to hallucination, and significantly impair bug isolation performance. Exploring methods that eliminate reliance on prompt engineering and instead allow the LLM to automatically request necessary information is promising. This approach offers the LLM greater flexibility, enabling it to actively acquire necessary supplementary data on demand during bug isolation. It reduces human intervention and can potentially further enhance automation and performance in complex scenarios.
\section{Related Work}

In this section, we discuss the research most relevant to our work. As mentioned in the introduction, existing compiler bug isolation techniques primarily fall into two categories: mutating test programs and mutating compilation configurations. In the first category, Chen et al.~\cite{DiWi} introduced DiWi, which generates a set of passing test programs by applying mutation operations to the failing test program. Subsequently, Chen et al.~\cite{RecBi} developed RecBi, which enhances DiWi by incorporating structural mutations and using reinforcement learning to guide the program generation process. Recently, Tu et al.~\cite{llm4cbi} introduced LLM4CBI, a method that leverages LLMs to guide the generation of passing test programs for compiler bug isolation.
For the second category, Yang et al.~\cite{yang2022isolating} introduced ODFL, which isolates compiler bugs by analyzing coverage differences of the failing test program across different compilation configurations (e.g., O2 vs O3). Similarly, Zhou et al.~\cite{zhou2022locseq} proposed LocSeq, which targets compiler bugs caused by interactions in optimization sequences. It constructs optimization sequences that are similar to the bug-triggering sequence but do not trigger the bug, using coverage differences to identify buggy locations. 
Additionally, Zeller~\cite{zeller2002isolating} introduced delta debugging, which isolates compiler bugs by examining runtime program states between passing and failing runs. Holmes and Groce~\cite{holmes2018causal,holmes2020using} proposed using mutation testing to locate compiler bugs; if two failing test programs for a compiler can be made to pass by applying the same mutation, they are likely due to the same fault. Chang et al.~\cite{chang2005type} suggested utilizing specially-designed type-based data-flow analysis to aid compiler bug debugging, while Lim and Debray~\cite{lim2021automated} employed dynamic analysis techniques to identify errors in JIT compilers. Yang et al.~\cite{llmao} proposed LLMAO, which leverages LLMs for localizing program bugs by taking source code as input while ignoring other features.


Unlike these approaches, our method seeks to enhance compiler bug isolation performance by leveraging LLMs to integrate multiple data sources, rather than relying solely on individual information such as test coverage or program states. Although the latest FuseFL~\cite{fusefl} also leverages LLMs to combine multiple contextual features, the used features and method are still largely different. Our experimental results demonstrate that our approach significantly outperformed FuseFL.
In fact, our approach complements these existing methods and can potentially be integrated with them by taking their results as extra input for additional improvements.

\section{Conclusion}
\label{sec:conclusion}
In this paper, we introduce \tool{}, a novel compiler bug isolation method that leverages the power of LLMs to integrate multiple data sources, including failing test programs, source file summaries, lists of suspicious files identified through both coarse-grained and finer-grained test execution coverage, as well as compilation configurations with related output messages.
This comprehensive contextual information guides the LLM in effectively localizing defects in compilers. 
Our experimental results using a widely-used benchmark dataset demonstrate that \tool{} successfully isolated 42/63/78/95 out of 120 bugs within Top-1/5/10/20 positions.
Remarkably, \tool{} outperformed the best-performing RecBi~\cite{RecBi} by isolating 66.67\% and 26.92\% more bugs regarding Top-1 and Top-5 accuracies for GCC, and 69.23\% and 15.38\% more bugs for LLVM. 
Furthermore, the results highlight the robustness of \tool{}, which consistently maintains high effectiveness across compilers and configurations, 
surpassing existing advanced techniques.

\bibliographystyle{ACM-Reference-Format}
\bibliography{references}


\begin{thebibliography}{51}


\ifx \showCODEN    \undefined \def \showCODEN     #1{\unskip}     \fi
\ifx \showISBNx    \undefined \def \showISBNx     #1{\unskip}     \fi
\ifx \showISBNxiii \undefined \def \showISBNxiii  #1{\unskip}     \fi
\ifx \showISSN     \undefined \def \showISSN      #1{\unskip}     \fi
\ifx \showLCCN     \undefined \def \showLCCN      #1{\unskip}     \fi
\ifx \shownote     \undefined \def \shownote      #1{#1}          \fi
\ifx \showarticletitle \undefined \def \showarticletitle #1{#1}   \fi
\ifx \showURL      \undefined \def \showURL       {\relax}        \fi
\providecommand\bibfield[2]{#2}
\providecommand\bibinfo[2]{#2}
\providecommand\natexlab[1]{#1}
\providecommand\showeprint[2][]{arXiv:#2}

\bibitem[GCC(2013)]%
        {GCCBUGID59221}
 \bibinfo{year}{2013}\natexlab{}.
\newblock \bibinfo{title}{{GCC BUG ID 59221}}.
\newblock \bibinfo{howpublished}{\url{https://gcc.gnu.org/bugzilla/show_bug.cgi?id=59221}}.
\newblock
\newblock
\shownote{Accessed: 2025}.


\bibitem[LLV(2015)]%
        {LLVMBUGID25154}
 \bibinfo{year}{2015}\natexlab{}.
\newblock \bibinfo{title}{{LLVM BUG ID 25154}}.
\newblock \bibinfo{howpublished}{\url{https://bugs.llvm.org/show_bug.cgi?id=25154}}.
\newblock
\newblock
\shownote{Accessed: 2025}.


\bibitem[Abreu et~al\mbox{.}(2007a)]%
        {abreu2007accuracy}
\bibfield{author}{\bibinfo{person}{Rui Abreu}, \bibinfo{person}{Peter Zoeteweij}, {and} \bibinfo{person}{Arjan~JC Van~Gemund}.} \bibinfo{year}{2007}\natexlab{a}.
\newblock \showarticletitle{On the accuracy of spectrum-based fault localization}. In \bibinfo{booktitle}{\emph{Testing: Academic and industrial conference practice and research techniques-MUTATION (TAICPART-MUTATION 2007)}}. IEEE, \bibinfo{pages}{89--98}.
\newblock


\bibitem[Abreu et~al\mbox{.}(2007b)]%
        {ochiai}
\bibfield{author}{\bibinfo{person}{Rui Abreu}, \bibinfo{person}{Peter Zoeteweij}, {and} \bibinfo{person}{Arjan~JC Van~Gemund}.} \bibinfo{year}{2007}\natexlab{b}.
\newblock \showarticletitle{On the accuracy of spectrum-based fault localization}. In \bibinfo{booktitle}{\emph{Testing: Academic and industrial conference practice and research techniques-MUTATION (TAICPART-MUTATION 2007)}}. IEEE, \bibinfo{pages}{89--98}.
\newblock


\bibitem[Abreu et~al\mbox{.}(2009)]%
        {barinel}
\bibfield{author}{\bibinfo{person}{Rui Abreu}, \bibinfo{person}{Peter Zoeteweij}, {and} \bibinfo{person}{Arjan~JC Van~Gemund}.} \bibinfo{year}{2009}\natexlab{}.
\newblock \showarticletitle{Spectrum-based multiple fault localization}. In \bibinfo{booktitle}{\emph{2009 IEEE/ACM International Conference on Automated Software Engineering}}. IEEE, \bibinfo{pages}{88--99}.
\newblock


\bibitem[Benton et~al\mbox{.}(2020)]%
        {benton2020effectiveness}
\bibfield{author}{\bibinfo{person}{Samuel Benton}, \bibinfo{person}{Xia Li}, \bibinfo{person}{Yiling Lou}, {and} \bibinfo{person}{Lingming Zhang}.} \bibinfo{year}{2020}\natexlab{}.
\newblock \showarticletitle{On the effectiveness of unified debugging: An extensive study on 16 program repair systems}. In \bibinfo{booktitle}{\emph{Proceedings of the 35th IEEE/ACM International Conference on Automated Software Engineering}}. \bibinfo{pages}{907--918}.
\newblock


\bibitem[Chang et~al\mbox{.}(2005)]%
        {chang2005type}
\bibfield{author}{\bibinfo{person}{Bor-Yuh~Evan Chang}, \bibinfo{person}{Adam Chlipala}, \bibinfo{person}{George~C Necula}, {and} \bibinfo{person}{Robert~R Schneck}.} \bibinfo{year}{2005}\natexlab{}.
\newblock \showarticletitle{Type-based verification of sssembly language for compiler debugging}. In \bibinfo{booktitle}{\emph{Proceedings of the 2005 ACM SIGPLAN international workshop on Types in languages design and implementation}}. \bibinfo{pages}{91--102}.
\newblock


\bibitem[Chen et~al\mbox{.}(2019)]%
        {DiWi}
\bibfield{author}{\bibinfo{person}{Junjie Chen}, \bibinfo{person}{Jiaqi Han}, \bibinfo{person}{Peiyi Sun}, \bibinfo{person}{Lingming Zhang}, \bibinfo{person}{Dan Hao}, {and} \bibinfo{person}{Lu Zhang}.} \bibinfo{year}{2019}\natexlab{}.
\newblock \showarticletitle{Compiler bug isolation via effective witness test program generation}. In \bibinfo{booktitle}{\emph{Proceedings of the 2019 27th ACM Joint Meeting on European Software Engineering Conference and Symposium on the Foundations of Software Engineering}}. \bibinfo{pages}{223--234}.
\newblock


\bibitem[Chen et~al\mbox{.}(2020a)]%
        {RecBi}
\bibfield{author}{\bibinfo{person}{Junjie Chen}, \bibinfo{person}{Haoyang Ma}, {and} \bibinfo{person}{Lingming Zhang}.} \bibinfo{year}{2020}\natexlab{a}.
\newblock \showarticletitle{Enhanced compiler bug isolation via memoized search}. In \bibinfo{booktitle}{\emph{Proceedings of the 35th IEEE/ACM International Conference on Automated Software Engineering}}. \bibinfo{pages}{78--89}.
\newblock


\bibitem[Chen et~al\mbox{.}(2020b)]%
        {chen2020survey}
\bibfield{author}{\bibinfo{person}{Junjie Chen}, \bibinfo{person}{Jibesh Patra}, \bibinfo{person}{Michael Pradel}, \bibinfo{person}{Yingfei Xiong}, \bibinfo{person}{Hongyu Zhang}, \bibinfo{person}{Dan Hao}, {and} \bibinfo{person}{Lu Zhang}.} \bibinfo{year}{2020}\natexlab{b}.
\newblock \showarticletitle{A survey of compiler testing}.
\newblock \bibinfo{journal}{\emph{ACM Computing Surveys (CSUR)}} \bibinfo{volume}{53}, \bibinfo{number}{1} (\bibinfo{year}{2020}), \bibinfo{pages}{1--36}.
\newblock


\bibitem[{DeepSeek}(2024)]%
        {DeepSeek-V3}
\bibfield{author}{\bibinfo{person}{{DeepSeek}}.} \bibinfo{year}{2024}\natexlab{}.
\newblock \bibinfo{title}{{DeepSeek-V3-671B}}.
\newblock \bibinfo{howpublished}{\url{https://api-docs.deepseek.com/zh-cn/news/news1226}}.
\newblock
\newblock
\shownote{Accessed: 2025}.


\bibitem[{DeepSeek}(2025)]%
        {DeepSeek-R1}
\bibfield{author}{\bibinfo{person}{{DeepSeek}}.} \bibinfo{year}{2025}\natexlab{}.
\newblock \bibinfo{title}{{DeepSeek-R1-671B}}.
\newblock \bibinfo{howpublished}{\url{https://api-docs.deepseek.com/zh-cn/news/news250120}}.
\newblock
\newblock
\shownote{Accessed: 2025}.


\bibitem[DiGiuseppe and Jones(2011)]%
        {digiuseppe2011influence}
\bibfield{author}{\bibinfo{person}{Nicholas DiGiuseppe} {and} \bibinfo{person}{James~A Jones}.} \bibinfo{year}{2011}\natexlab{}.
\newblock \showarticletitle{On the influence of multiple faults on coverage-based fault localization}. In \bibinfo{booktitle}{\emph{Proceedings of the 2011 international symposium on software testing and analysis}}. \bibinfo{pages}{210--220}.
\newblock


\bibitem[D'Silva et~al\mbox{.}(2015)]%
        {d2015correctness}
\bibfield{author}{\bibinfo{person}{Vijay D'Silva}, \bibinfo{person}{Mathias Payer}, {and} \bibinfo{person}{Dawn Song}.} \bibinfo{year}{2015}\natexlab{}.
\newblock \showarticletitle{The correctness-security gap in compiler optimization}. In \bibinfo{booktitle}{\emph{2015 IEEE Security and Privacy Workshops}}. IEEE, \bibinfo{pages}{73--87}.
\newblock


\bibitem[Duran and Ntafos(1984)]%
        {duran1984evaluation}
\bibfield{author}{\bibinfo{person}{Joe~W Duran} {and} \bibinfo{person}{Simeon~C Ntafos}.} \bibinfo{year}{1984}\natexlab{}.
\newblock \showarticletitle{An evaluation of random testing}.
\newblock \bibinfo{journal}{\emph{IEEE transactions on Software Engineering}} \bibinfo{number}{4} (\bibinfo{year}{1984}), \bibinfo{pages}{438--444}.
\newblock


\bibitem[{GCC}({[n.\,d.]})]%
        {gcc}
\bibfield{author}{\bibinfo{person}{{GCC}}.} \bibinfo{year}{[n.\,d.]}\natexlab{}.
\newblock \bibinfo{title}{{GCC}}.
\newblock \bibinfo{howpublished}{\url{https://gcc.gnu.org}}.
\newblock
\newblock
\shownote{Accessed: 2024}.


\bibitem[Godefroid et~al\mbox{.}(2008)]%
        {godefroid2008automated}
\bibfield{author}{\bibinfo{person}{Patrice Godefroid}, \bibinfo{person}{Michael~Y Levin}, \bibinfo{person}{David~A Molnar}, {et~al\mbox{.}}} \bibinfo{year}{2008}\natexlab{}.
\newblock \showarticletitle{Automated whitebox fuzz testing.}. In \bibinfo{booktitle}{\emph{NDSS}}, Vol.~\bibinfo{volume}{8}. \bibinfo{pages}{151--166}.
\newblock


\bibitem[Holmes and Groce(2018)]%
        {holmes2018causal}
\bibfield{author}{\bibinfo{person}{Josie Holmes} {and} \bibinfo{person}{Alex Groce}.} \bibinfo{year}{2018}\natexlab{}.
\newblock \showarticletitle{Causal distance-metric-based assistance for debugging after compiler fuzzing}. In \bibinfo{booktitle}{\emph{2018 IEEE 29th International Symposium on Software Reliability Engineering (ISSRE)}}. IEEE, \bibinfo{pages}{166--177}.
\newblock


\bibitem[Holmes and Groce(2020)]%
        {holmes2020using}
\bibfield{author}{\bibinfo{person}{Josie Holmes} {and} \bibinfo{person}{Alex Groce}.} \bibinfo{year}{2020}\natexlab{}.
\newblock \showarticletitle{Using mutants to help developers distinguish and debug (compiler) faults}.
\newblock \bibinfo{journal}{\emph{Software Testing, Verification and Reliability}} \bibinfo{volume}{30}, \bibinfo{number}{2} (\bibinfo{year}{2020}), \bibinfo{pages}{e1727}.
\newblock


\bibitem[Hou et~al\mbox{.}(2023)]%
        {hou2023large}
\bibfield{author}{\bibinfo{person}{Xinyi Hou}, \bibinfo{person}{Yanjie Zhao}, \bibinfo{person}{Yue Liu}, \bibinfo{person}{Zhou Yang}, \bibinfo{person}{Kailong Wang}, \bibinfo{person}{Li Li}, \bibinfo{person}{Xiapu Luo}, \bibinfo{person}{David Lo}, \bibinfo{person}{John Grundy}, {and} \bibinfo{person}{Haoyu Wang}.} \bibinfo{year}{2023}\natexlab{}.
\newblock \showarticletitle{Large language models for software engineering: A systematic literature review}.
\newblock \bibinfo{journal}{\emph{ACM Transactions on Software Engineering and Methodology}} (\bibinfo{year}{2023}).
\newblock


\bibitem[Jiang et~al\mbox{.}(2019)]%
        {jiang2019manual}
\bibfield{author}{\bibinfo{person}{Jiajun Jiang}, \bibinfo{person}{Yingfei Xiong}, {and} \bibinfo{person}{Xin Xia}.} \bibinfo{year}{2019}\natexlab{}.
\newblock \showarticletitle{A manual inspection of Defects4J bugs and its implications for automatic program repair}.
\newblock \bibinfo{journal}{\emph{Science china information sciences}}  \bibinfo{volume}{62} (\bibinfo{year}{2019}), \bibinfo{pages}{1--16}.
\newblock


\bibitem[Jin et~al\mbox{.}(2023)]%
        {jin2023inferfix}
\bibfield{author}{\bibinfo{person}{Matthew Jin}, \bibinfo{person}{Syed Shahriar}, \bibinfo{person}{Michele Tufano}, \bibinfo{person}{Xin Shi}, \bibinfo{person}{Shuai Lu}, \bibinfo{person}{Neel Sundaresan}, {and} \bibinfo{person}{Alexey Svyatkovskiy}.} \bibinfo{year}{2023}\natexlab{}.
\newblock \showarticletitle{Inferfix: End-to-end program repair with llms}. In \bibinfo{booktitle}{\emph{Proceedings of the 31st ACM Joint European Software Engineering Conference and Symposium on the Foundations of Software Engineering}}. \bibinfo{pages}{1646--1656}.
\newblock


\bibitem[Jones and Harrold(2005)]%
        {tarantula}
\bibfield{author}{\bibinfo{person}{James~A Jones} {and} \bibinfo{person}{Mary~Jean Harrold}.} \bibinfo{year}{2005}\natexlab{}.
\newblock \showarticletitle{Empirical evaluation of the tarantula automatic fault-localization technique}. In \bibinfo{booktitle}{\emph{Proceedings of the 20th IEEE/ACM international Conference on Automated software engineering}}. \bibinfo{pages}{273--282}.
\newblock


\bibitem[Kochhar et~al\mbox{.}(2016)]%
        {practitioners2016Kochhar}
\bibfield{author}{\bibinfo{person}{Pavneet~Singh Kochhar}, \bibinfo{person}{Xin Xia}, \bibinfo{person}{David Lo}, {and} \bibinfo{person}{Shanping Li}.} \bibinfo{year}{2016}\natexlab{}.
\newblock \showarticletitle{Practitioners' expectations on automated fault localization}. In \bibinfo{booktitle}{\emph{Proceedings of the 25th International Symposium on Software Testing and Analysis}} (Saarbr\"{u}cken, Germany) \emph{(\bibinfo{series}{ISSTA 2016})}. \bibinfo{publisher}{Association for Computing Machinery}, \bibinfo{address}{New York, NY, USA}, \bibinfo{pages}{165–176}.
\newblock
\showISBNx{9781450343909}
\href{https://doi.org/10.1145/2931037.2931051}{doi:\nolinkurl{10.1145/2931037.2931051}}


\bibitem[LeCun et~al\mbox{.}(2015)]%
        {lecun2015deep}
\bibfield{author}{\bibinfo{person}{Yann LeCun}, \bibinfo{person}{Yoshua Bengio}, {and} \bibinfo{person}{Geoffrey Hinton}.} \bibinfo{year}{2015}\natexlab{}.
\newblock \showarticletitle{Deep learning}.
\newblock \bibinfo{journal}{\emph{nature}} \bibinfo{volume}{521}, \bibinfo{number}{7553} (\bibinfo{year}{2015}), \bibinfo{pages}{436--444}.
\newblock


\bibitem[Li and Zhang(2017)]%
        {li2017transforming}
\bibfield{author}{\bibinfo{person}{Xia Li} {and} \bibinfo{person}{Lingming Zhang}.} \bibinfo{year}{2017}\natexlab{}.
\newblock \showarticletitle{Transforming programs and tests in tandem for fault localization}.
\newblock \bibinfo{journal}{\emph{Proceedings of the ACM on Programming Languages}} \bibinfo{volume}{1}, \bibinfo{number}{OOPSLA} (\bibinfo{year}{2017}), \bibinfo{pages}{1--30}.
\newblock


\bibitem[Lim and Debray(2021)]%
        {lim2021automated}
\bibfield{author}{\bibinfo{person}{HeuiChan Lim} {and} \bibinfo{person}{Saumya Debray}.} \bibinfo{year}{2021}\natexlab{}.
\newblock \showarticletitle{Automated bug localization in JIT compilers}. In \bibinfo{booktitle}{\emph{Proceedings of the 17th ACM SIGPLAN/SIGOPS International Conference on Virtual Execution Environments}}. \bibinfo{pages}{153--164}.
\newblock


\bibitem[{LLVM}({[n.\,d.]})]%
        {llvm}
\bibfield{author}{\bibinfo{person}{{LLVM}}.} \bibinfo{year}{[n.\,d.]}\natexlab{}.
\newblock \bibinfo{title}{{LLVM}}.
\newblock \bibinfo{howpublished}{\url{https://llvm.org}}.
\newblock
\newblock
\shownote{Accessed: 2024}.


\bibitem[Lou et~al\mbox{.}(2020)]%
        {lou2020can}
\bibfield{author}{\bibinfo{person}{Yiling Lou}, \bibinfo{person}{Ali Ghanbari}, \bibinfo{person}{Xia Li}, \bibinfo{person}{Lingming Zhang}, \bibinfo{person}{Haotian Zhang}, \bibinfo{person}{Dan Hao}, {and} \bibinfo{person}{Lu Zhang}.} \bibinfo{year}{2020}\natexlab{}.
\newblock \showarticletitle{Can automated program repair refine fault localization? a unified debugging approach}. In \bibinfo{booktitle}{\emph{Proceedings of the 29th ACM SIGSOFT International Symposium on Software Testing and Analysis}}. \bibinfo{pages}{75--87}.
\newblock


\bibitem[Min et~al\mbox{.}(2023)]%
        {min2023recent}
\bibfield{author}{\bibinfo{person}{Bonan Min}, \bibinfo{person}{Hayley Ross}, \bibinfo{person}{Elior Sulem}, \bibinfo{person}{Amir Pouran~Ben Veyseh}, \bibinfo{person}{Thien~Huu Nguyen}, \bibinfo{person}{Oscar Sainz}, \bibinfo{person}{Eneko Agirre}, \bibinfo{person}{Ilana Heintz}, {and} \bibinfo{person}{Dan Roth}.} \bibinfo{year}{2023}\natexlab{}.
\newblock \showarticletitle{Recent advances in natural language processing via large pre-trained language models: A survey}.
\newblock \bibinfo{journal}{\emph{Comput. Surveys}} \bibinfo{volume}{56}, \bibinfo{number}{2} (\bibinfo{year}{2023}), \bibinfo{pages}{1--40}.
\newblock


\bibitem[{OpenAI}(2024)]%
        {ChatGPT-4o}
\bibfield{author}{\bibinfo{person}{{OpenAI}}.} \bibinfo{year}{2024}\natexlab{}.
\newblock \bibinfo{title}{{ChatGPT-4o-2024-08-06}}.
\newblock \bibinfo{howpublished}{\url{https://platform.openai.com/docs/models\#gpt-4o}}.
\newblock
\newblock
\shownote{Accessed: 2025}.


\bibitem[Romano et~al\mbox{.}(2021)]%
        {romano2021empirical}
\bibfield{author}{\bibinfo{person}{Alan Romano}, \bibinfo{person}{Xinyue Liu}, \bibinfo{person}{Yonghwi Kwon}, {and} \bibinfo{person}{Weihang Wang}.} \bibinfo{year}{2021}\natexlab{}.
\newblock \showarticletitle{An empirical study of bugs in webassembly compilers}. In \bibinfo{booktitle}{\emph{2021 36th IEEE/ACM International Conference on Automated Software Engineering (ASE)}}. IEEE, \bibinfo{pages}{42--54}.
\newblock


\bibitem[Santelices et~al\mbox{.}(2009)]%
        {sbfl4}
\bibfield{author}{\bibinfo{person}{Raul Santelices}, \bibinfo{person}{James~A Jones}, \bibinfo{person}{Yanbing Yu}, {and} \bibinfo{person}{Mary~Jean Harrold}.} \bibinfo{year}{2009}\natexlab{}.
\newblock \showarticletitle{Lightweight fault-localization using multiple coverage types}. In \bibinfo{booktitle}{\emph{2009 IEEE 31st International Conference on Software Engineering}}. IEEE, \bibinfo{pages}{56--66}.
\newblock


\bibitem[Sidhpurwala(2019)]%
        {sidhpurwala2019security}
\bibfield{author}{\bibinfo{person}{Huzaifa Sidhpurwala}.} \bibinfo{year}{2019}\natexlab{}.
\newblock \showarticletitle{Security flaws caused by compiler optimizations}.
\newblock \bibinfo{journal}{\emph{Red Hat Blog}} (\bibinfo{year}{2019}).
\newblock


\bibitem[Sun et~al\mbox{.}(2016)]%
        {sun2016finding}
\bibfield{author}{\bibinfo{person}{Chengnian Sun}, \bibinfo{person}{Vu Le}, {and} \bibinfo{person}{Zhendong Su}.} \bibinfo{year}{2016}\natexlab{}.
\newblock \showarticletitle{Finding and analyzing compiler warning defects}. In \bibinfo{booktitle}{\emph{Proceedings of the 38th International Conference on Software Engineering}}. \bibinfo{pages}{203--213}.
\newblock


\bibitem[Tu et~al\mbox{.}(2024)]%
        {llm4cbi}
\bibfield{author}{\bibinfo{person}{Haoxin Tu}, \bibinfo{person}{Zhide Zhou}, \bibinfo{person}{He Jiang}, \bibinfo{person}{Imam Nur~Bani Yusuf}, \bibinfo{person}{Yuxian Li}, {and} \bibinfo{person}{Lingxiao Jiang}.} \bibinfo{year}{2024}\natexlab{}.
\newblock \showarticletitle{Isolating Compiler Bugs by Generating Effective Witness Programs with Large Language Models}.
\newblock \bibinfo{journal}{\emph{IEEE Transactions on Software Engineering}} (\bibinfo{year}{2024}).
\newblock


\bibitem[Wang et~al\mbox{.}(2015)]%
        {wang2015evaluating}
\bibfield{author}{\bibinfo{person}{Qianqian Wang}, \bibinfo{person}{Chris Parnin}, {and} \bibinfo{person}{Alessandro Orso}.} \bibinfo{year}{2015}\natexlab{}.
\newblock \showarticletitle{Evaluating the usefulness of ir-based fault localization techniques}. In \bibinfo{booktitle}{\emph{Proceedings of the 2015 international symposium on software testing and analysis}}. \bibinfo{pages}{1--11}.
\newblock


\bibitem[Wang et~al\mbox{.}(2013)]%
        {wang2013towards}
\bibfield{author}{\bibinfo{person}{Xi Wang}, \bibinfo{person}{Nickolai Zeldovich}, \bibinfo{person}{M~Frans Kaashoek}, {and} \bibinfo{person}{Armando Solar-Lezama}.} \bibinfo{year}{2013}\natexlab{}.
\newblock \showarticletitle{Towards optimization-safe systems: Analyzing the impact of undefined behavior}. In \bibinfo{booktitle}{\emph{Proceedings of the Twenty-Fourth ACM Symposium on Operating Systems Principles}}. \bibinfo{pages}{260--275}.
\newblock


\bibitem[Wen et~al\mbox{.}(2019)]%
        {wen2019historical}
\bibfield{author}{\bibinfo{person}{Ming Wen}, \bibinfo{person}{Junjie Chen}, \bibinfo{person}{Yongqiang Tian}, \bibinfo{person}{Rongxin Wu}, \bibinfo{person}{Dan Hao}, \bibinfo{person}{Shi Han}, {and} \bibinfo{person}{Shing-Chi Cheung}.} \bibinfo{year}{2019}\natexlab{}.
\newblock \showarticletitle{Historical spectrum based fault localization}.
\newblock \bibinfo{journal}{\emph{IEEE Transactions on Software Engineering}} \bibinfo{volume}{47}, \bibinfo{number}{11} (\bibinfo{year}{2019}), \bibinfo{pages}{2348--2368}.
\newblock


\bibitem[Widyasari et~al\mbox{.}(2024)]%
        {fusefl}
\bibfield{author}{\bibinfo{person}{Ratnadira Widyasari}, \bibinfo{person}{Jia~Wei Ang}, \bibinfo{person}{Truong~Giang Nguyen}, \bibinfo{person}{Neil Sharma}, {and} \bibinfo{person}{David Lo}.} \bibinfo{year}{2024}\natexlab{}.
\newblock \showarticletitle{Demystifying faulty code: Step-by-step reasoning for explainable fault localization}. In \bibinfo{booktitle}{\emph{2024 IEEE International Conference on Software Analysis, Evolution and Reengineering (SANER)}}. IEEE, \bibinfo{pages}{568--579}.
\newblock


\bibitem[Wong et~al\mbox{.}(2013)]%
        {Dstar}
\bibfield{author}{\bibinfo{person}{W~Eric Wong}, \bibinfo{person}{Vidroha Debroy}, \bibinfo{person}{Ruizhi Gao}, {and} \bibinfo{person}{Yihao Li}.} \bibinfo{year}{2013}\natexlab{}.
\newblock \showarticletitle{The DStar method for effective software fault localization}.
\newblock \bibinfo{journal}{\emph{IEEE Transactions on Reliability}} \bibinfo{volume}{63}, \bibinfo{number}{1} (\bibinfo{year}{2013}), \bibinfo{pages}{290--308}.
\newblock


\bibitem[Wong et~al\mbox{.}(2007)]%
        {wong2}
\bibfield{author}{\bibinfo{person}{W~Eric Wong}, \bibinfo{person}{Yu Qi}, \bibinfo{person}{Lei Zhao}, {and} \bibinfo{person}{Kai-Yuan Cai}.} \bibinfo{year}{2007}\natexlab{}.
\newblock \showarticletitle{Effective fault localization using code coverage}. In \bibinfo{booktitle}{\emph{31st Annual International Computer Software and Applications Conference (COMPSAC 2007)}}, Vol.~\bibinfo{volume}{1}. IEEE, \bibinfo{pages}{449--456}.
\newblock


\bibitem[Xuan and Monperrus(2014)]%
        {xuan2014test}
\bibfield{author}{\bibinfo{person}{Jifeng Xuan} {and} \bibinfo{person}{Martin Monperrus}.} \bibinfo{year}{2014}\natexlab{}.
\newblock \showarticletitle{Test case purification for improving fault localization}. In \bibinfo{booktitle}{\emph{Proceedings of the 22nd ACM SIGSOFT international symposium on foundations of software engineering}}. \bibinfo{pages}{52--63}.
\newblock


\bibitem[Yang et~al\mbox{.}(2024)]%
        {llmao}
\bibfield{author}{\bibinfo{person}{Aidan~ZH Yang}, \bibinfo{person}{Claire Le~Goues}, \bibinfo{person}{Ruben Martins}, {and} \bibinfo{person}{Vincent Hellendoorn}.} \bibinfo{year}{2024}\natexlab{}.
\newblock \showarticletitle{Large language models for test-free fault localization}. In \bibinfo{booktitle}{\emph{Proceedings of the 46th IEEE/ACM International Conference on Software Engineering}}. \bibinfo{pages}{1--12}.
\newblock


\bibitem[Yang et~al\mbox{.}(2022)]%
        {yang2022isolating}
\bibfield{author}{\bibinfo{person}{Jing Yang}, \bibinfo{person}{Yibiao Yang}, \bibinfo{person}{Maolin Sun}, \bibinfo{person}{Ming Wen}, \bibinfo{person}{Yuming Zhou}, {and} \bibinfo{person}{Hai Jin}.} \bibinfo{year}{2022}\natexlab{}.
\newblock \showarticletitle{Isolating compiler optimization faults via differentiating finer-grained options}. In \bibinfo{booktitle}{\emph{2022 IEEE International Conference on Software Analysis, Evolution and Reengineering (SANER)}}. IEEE, \bibinfo{pages}{481--491}.
\newblock


\bibitem[Zeller(2002)]%
        {zeller2002isolating}
\bibfield{author}{\bibinfo{person}{Andreas Zeller}.} \bibinfo{year}{2002}\natexlab{}.
\newblock \showarticletitle{Isolating cause-effect chains from computer programs}.
\newblock \bibinfo{journal}{\emph{ACM SIGSOFT Software Engineering Notes}} \bibinfo{volume}{27}, \bibinfo{number}{6} (\bibinfo{year}{2002}), \bibinfo{pages}{1--10}.
\newblock


\bibitem[Zhang et~al\mbox{.}(2011)]%
        {zhang2011localizing}
\bibfield{author}{\bibinfo{person}{Lingming Zhang}, \bibinfo{person}{Miryung Kim}, {and} \bibinfo{person}{Sarfraz Khurshid}.} \bibinfo{year}{2011}\natexlab{}.
\newblock \showarticletitle{Localizing failure-inducing program edits based on spectrum information}. In \bibinfo{booktitle}{\emph{2011 27th IEEE International Conference on Software Maintenance (ICSM)}}. IEEE, \bibinfo{pages}{23--32}.
\newblock


\bibitem[Zhang et~al\mbox{.}(2013)]%
        {zhang2013injecting}
\bibfield{author}{\bibinfo{person}{Lingming Zhang}, \bibinfo{person}{Lu Zhang}, {and} \bibinfo{person}{Sarfraz Khurshid}.} \bibinfo{year}{2013}\natexlab{}.
\newblock \showarticletitle{Injecting mechanical faults to localize developer faults for evolving software}.
\newblock \bibinfo{journal}{\emph{ACM SIGPLAN Notices}} \bibinfo{volume}{48}, \bibinfo{number}{10} (\bibinfo{year}{2013}), \bibinfo{pages}{765--784}.
\newblock


\bibitem[Zhang et~al\mbox{.}(2024)]%
        {zhang2024autocoderover}
\bibfield{author}{\bibinfo{person}{Yuntong Zhang}, \bibinfo{person}{Haifeng Ruan}, \bibinfo{person}{Zhiyu Fan}, {and} \bibinfo{person}{Abhik Roychoudhury}.} \bibinfo{year}{2024}\natexlab{}.
\newblock \showarticletitle{Autocoderover: Autonomous program improvement}. In \bibinfo{booktitle}{\emph{Proceedings of the 33rd ACM SIGSOFT International Symposium on Software Testing and Analysis}}. \bibinfo{pages}{1592--1604}.
\newblock


\bibitem[Zhou et~al\mbox{.}(2012)]%
        {6227210}
\bibfield{author}{\bibinfo{person}{Jian Zhou}, \bibinfo{person}{Hongyu Zhang}, {and} \bibinfo{person}{David Lo}.} \bibinfo{year}{2012}\natexlab{}.
\newblock \showarticletitle{Where should the bugs be fixed? More accurate information retrieval-based bug localization based on bug reports}. In \bibinfo{booktitle}{\emph{2012 34th International Conference on Software Engineering (ICSE)}}. \bibinfo{pages}{14--24}.
\newblock
\href{https://doi.org/10.1109/ICSE.2012.6227210}{doi:\nolinkurl{10.1109/ICSE.2012.6227210}}


\bibitem[Zhou et~al\mbox{.}(2022)]%
        {zhou2022locseq}
\bibfield{author}{\bibinfo{person}{Zhide Zhou}, \bibinfo{person}{He Jiang}, \bibinfo{person}{Zhilei Ren}, \bibinfo{person}{Yuting Chen}, {and} \bibinfo{person}{Lei Qiao}.} \bibinfo{year}{2022}\natexlab{}.
\newblock \showarticletitle{Locseq: Automated localization for compiler optimization sequence bugs of LLVM}.
\newblock \bibinfo{journal}{\emph{IEEE Transactions on Reliability}} \bibinfo{volume}{71}, \bibinfo{number}{2} (\bibinfo{year}{2022}), \bibinfo{pages}{896--910}.
\newblock


\end{thebibliography}
\end{document}